\documentclass[11pt,titlepage,a4paper]{myarticle}

\usepackage[T1]{fontenc} 
\RequirePackage[colorlinks=true
,urlcolor=blue
,anchorcolor=blue
,citecolor=blue
,filecolor=blue
,linkcolor=blue
,menucolor=blue
,linktocpage=true
,pdfproducer=medialab
,pdfa=true
]{hyperref}
\usepackage{amsmath, mathrsfs, amsfonts, amssymb, amsthm, mathtools, graphicx, color, ucs, xparse, tikz, lmodern, physics, varioref, cleveref, tensor, cite}

\crefname{pluralequation}{eqs.}{eqs.}
\Crefname{pluralequation}{Eqs.}{Eqs.}
\crefformat{pluralequation}{eqs.~(#2#1#3)}
\Crefformat{pluralequation}{Eqs.~(#2#1#3)}

\renewcommand{\d}{\partial}

\DeclareMathOperator\arccosh{arccosh}

\newcommand{\overbar}[1]{\mkern 2mu\overline{\mkern-2mu#1\mkern-2mu}\mkern 2mu}
\newcommand{\C}{\mathbb{C}}

\newcommand{\Z}{\mathbb{Z}}

\newcommand{\hodge}{{\star}}

\newcommand\mperiod[1][\rlap]{#1{\,.}}	
\newcommand\mcomma[1][\rlap]{#1{\,,}}

\crefname{table}{table}{tables}
\Crefname{table}{Table}{Tables}
\crefname{figure}{figure}{figures}
\Crefname{figure}{Figure}{Figures}

\newenvironment{eq}
    {\begin{equation}
    \begin{aligned}
    }
    { 
    \end{aligned}
    \end{equation}
    \ignorespacesafterend
    }

\begin{document} 

\preprint{
	{\tt IFT-UAM/CSIC-25-101}\\
	}

\title{AdS vacua of non-supersymmetric strings}
\author{Salvatore Raucci
     \address{
     Instituto de F\'{i}sica Te\'{o}rica UAM/CSIC \\ and Departamento de F\'{i}sica Te\'{o}rica, \\
     Universidad Aut\'{o}noma de Madrid, Cantoblanco, 28049 Madrid, Spain\\ 
     {~}\\
} and Alessandro Tomasiello
    \address{
     Dipartimento di Matematica, \\ Universit\`{a} degli Studi di Milano–Bicocca, \\ Via Cozzi 55, 20126 Milano, Italy \\ and INFN, sezione di Milano–Bicocca\\
     {~}\\
     }
}

\Abstract{Few vacua are known for the three tachyon-free non-supersymmetric string theories. We find new classes of AdS backgrounds by focusing on spaces where the equations of motion reduce to purely algebraic conditions. Our first examples involve non-zero three-form fluxes supported either on direct product internal spaces or on $T_{p,q}$ geometries. For the ${\rm SO}(16)\times {\rm SO}(16)$ heterotic string, we then develop a method to engineer vacua with the addition of gauge fields. 
A formal Kaluza--Klein reduction yields complete solutions on a broad class of coset spaces $G/H$, automatically satisfying the three-form Bianchi identities with $H$-valued gauge fields.}

\maketitle
\setcounter{page}{1}

\tableofcontents

\section{Introduction}

The landscape of non-supersymmetric vacua in string theory remains largely unexplored.
Two issues are to blame: without spacetime supersymmetry, one loses both systematic methods to engineer vacua and a physical principle to argue for stability.
In this work, we focus on the former aspect. 

We shall consider specific non-supersymmetric models: ten-dimensional superstrings that are non-supersymmetric and free of tachyons in the spectrum. These are the heterotic ${\rm SO}(16)\times {\rm SO}(16)$ string~\cite{Ginsparg:1986wr,Dixon:1986iz}, type 0'B string theory~\cite{Sagnotti:1995ga,Sagnotti:1996qj}, and the Sugimoto orientifold of type IIB~\cite{Sugimoto:1999tx}; see~\cite{Leone:2025mwo} for a recent review. 
Their worldsheet description allows us to capture the effects of the absence of supersymmetry within the formalism of string perturbation theory. These effects enter the equations of motion for the massless string modes as string-loop corrections~\cite{Fischler:1986ci,Fischler:1986tb,Callan:1986bc,Tseytlin:1990mv}.
However, some of these are not corrections at all because their tree-level counterparts vanish.
One such term is the cosmological constant, which vanishes at tree level. For non-supersymmetric strings, the string-loop corrections to the cosmological constant term come from non-canceling worldsheet tadpoles and generate a scalar potential for the dilaton, the only uncharged massless scalar in the three models.
Finding flux vacua of non-supersymmetric strings means solving the equations with the additional deformation of this scalar \emph{tadpole} potential.

Lacking supergravity equations, the main approach has been to solve the system of coupled partial differential equations with sufficiently simple Freund--Rubin-like metrics~\cite{Gubser:2001zr,Mourad:2016xbk,Basile:2018irz,Antonelli:2019nar,Raucci:2022bjw,Baykara:2022cwj} or cohomogeneity-one backgrounds~\cite{Mourad:2021roa,Mourad:2024dur}. In this work, we first identify new AdS vacua by generalizing the former type. We then proceed to present a strategy for vacuum engineering that applies to the heterotic ${\rm SO}(16)\times {\rm SO}(16)$ string.
This is based on the properties of Kaluza--Klein reductions and is somewhat similar to that of~\cite{Duff:1985cm,garcia2024pluriclosed}.

In the supersymmetric case, one of the key early strategies for solving the supergravity equations was to reduce the equations of motion to algebraic conditions; this will be one of our ingredients.
To achieve this in the non-supersymmetric cases, one needs to somehow control both the Ricci and the stress-energy tensors. 
For the Ricci tensor, there are several options available, such as Einstein manifolds, products or fibrations involving them, and homogeneous manifolds. For the stress-energy tensor, we focus on solutions with constant dilaton profiles, so that $e^{\phi_0}$ acts as the small parameter granting control.\footnote{For open strings, this is not necessarily the case. See~\cite{Raucci:2022bjw} for an example where $e^{\phi_0}$ is small but higher-derivative corrections can become relevant.} We must then take care of the fluxes, in particular of the contractions $\iota_m F\cdot \iota_n F$. We know very little about these in general, with some exceptions when $F$ defines a $G$-structure, and again for homogeneous manifolds.

In the non-supersymmetric theories, we face an additional complication: fluxes are scarce and are usually harmonic forms. Neglecting the contribution of gauge fields, the three non-supersymmetric strings of interest have a harmonic three-form.
If we take this to be proportional to the real part of an $\mathrm{SU}(3)$-structure, we then need a complex manifold (not necessarily K\"ahler) with a trivial canonical bundle, and in particular $c_1=0$. At the same time, the Einstein equations require an Einstein metric with a positive curvature; therefore, this direction does not show promise. 

At first sight, homogeneous manifolds do not seem to fare much better: harmonic three-forms appear rarely \cite{lauret-will}, especially in low dimensions. We are thus led to include gauge fields.

With gauge fields, the universal three-form is no longer harmonic, and the techniques above to control the Ricci and the stress-energy tensors can succeed. This is the setup that we explore in this work, employing in particular homogeneous manifolds $G/H$ as internal spaces. We introduce gauge fields in the same way as they would enter Kaluza--Klein reductions; the gauge curvature $F$ takes values in $H$, which thus must be a subgroup of ${\rm SO}(16)\times {\rm SO}(16)$. 
This generates the three-form and the gauge fields from a fictitious higher-dimensional three-form, automatically solving the Bianchi identity $\dd H=-\frac12 {\rm tr} F\wedge F$ and some of the flux equations \cite{Duff:1985cm,garcia2024pluriclosed}. 
The Einstein equations become algebraic, but they must still be addressed separately. They only allow AdS vacua, but these do exist on several coset spaces. In particular, for AdS$_4$ we find that $\mathbb{F}(1,2;3)={\rm SU}(3)/{\rm U}(1)\times {\rm U}(1)$, $\mathbb{CP}^3= {\rm Sp}(2)/{\rm Sp}(1)\times {\rm U}(1)$, and $S^6=G_2/{\rm SU}(3)$ admit solutions that are under control, with gauge groups $H={\rm U}(1)\times {\rm U}(1)$, ${\rm Sp}(1)\times {\rm U}(1)$, and ${\rm SU}(3)$, respectively. 

Our solutions are not scale separated: the Kaluza--Klein and AdS scales are comparable. In fact, one can argue this for solutions with a product metric and constant dilaton, with the assumption that $m_\mathrm{KK}^2\sim R_{(d)}$, the internal curvature; we present this argument (similar to that of \cite{Gautason:2015tig}) in \cref{app:no_scale_separation}. Thus, in this respect, the tadpole term does not improve the situation relative to supersymmetric strings, just as it does not for the existence of de Sitter solutions \cite{Basile:2020mpt}.

This paper is organized as follows. \Cref{sec:3-form-vacua} contains the new vacua of the three non-supersymmetric ten-dimensional strings that we obtain with the universal three-form field strength. The heterotic string and the two orientifolds are treated separately because the supersymmetry-breaking term is different in the two cases.
Then, in \cref{sec:cpt-H-F}, we explain our Kaluza--Klein-inspired approach to include gauge fields by compactifying on homogeneous spaces. After schematically explaining the idea in \cref{sub:red}, we set our conventions and define the building blocks in \cref{sec:hom}. In \cref{ssec:coset-reduction}, we outline the general procedure. \Cref{ssec:ads-examples} contains explicit examples of AdS$_4$ vacua for the ${\rm SO}(16)\times {\rm SO}(16)$ string. We conclude with \cref{ssec:comments}, providing some observations on the formalism and on the open questions that remain.

\section{Three-form vacua: old and new}\label{sec:3-form-vacua}

For ten-dimensional non-supersymmetric strings, the leading terms in the two-derivative string-frame action for the massless modes are
\begin{eq}\label{eq:action}
    S=\frac{2\pi}{(2\pi \ell_s)^8}\int \sqrt{-g}\left[e^{-2\phi}\left(R+4(\d\phi)^2\right) - \sum_k \frac{1}{2}e^{\beta_k \phi}F_k^2 - V(\phi)\right] \mcomma
\end{eq}
with model-dependent values of $k$ and $\beta_k$, and with the scalar potential arising at one loop, $V(\phi)=T_1$, or at half loop, $V(\phi)=T_{\frac{1}{2}} e^{-\phi}$. We use the parameterization
\begin{eq}
    V(\phi)=T e^{\gamma\phi}\mcomma
\end{eq}
with model-dependent $\gamma$, to address both cases with unified notation.
Note that $T$ is positive for the three ten-dimensional tachyon-free strings. In \cref{eq:action} we have neglected the contributions from gauge fields. These will appear in \cref{sec:cpt-H-F}.

Demanding a constant dilaton profile, $\phi=\phi_0$, the equations of motion read
\begin{eq} \label[pluralequation]{eq:constant-dilaton}
    R_{MN} & = \sum_k \frac{1}{2}e^{(\beta_k+2)\phi_0}\iota_M F_k \cdot \iota_N F_k- \frac{k(\gamma+2)+\beta_k-\gamma}{8(2\gamma+5)} e^{(\beta_k+2)\phi_0} F_k^2 g_{MN} \mcomma \\
    T e^{\gamma\phi_0} & = \sum_k \frac{k-2\beta_k-5}{2(2\gamma+5)}e^{\beta_k\phi_0} F_k^2 \mperiod
\end{eq}

We now separately analyze the heterotic string with one-loop vacuum energy and the two orientifolds with half-loop tadpole potentials.

\subsection{One-loop heterotic}
\label{ssec:heterotic-vacua}

The only form field in the heterotic SO$(16)\times$SO$(16)$ string is the Kalb--Ramond $B_2$ with field strength $H_3$, which has $\beta_3=-2$.
\Cref{eq:constant-dilaton} become
\begin{eq}\label[pluralequation]{eq:heterotic-constant-dilaton}
    T e^{2\phi_0} & =\frac{1}{5} H_3^2 \mcomma \\
    R_{MN} & = \frac{1}{2}\iota_M H_3\cdot \iota_N H_3-\frac{1}{10}H_3^2 g_{MN}\mcomma
\end{eq}
and the positive sign of $T$ requires at least a magnetic $H_3$ flux. In fact, the simplest possibility is a spacetime of the form MS$_7\times X_3$, with a maximally symmetric seven-dimensional external spacetime MS$_7$, an Einstein manifold $X_3$, and with a three-form flux proportional to the internal volume form.
This solution is already known~\cite{Mourad:2016xbk,Basile:2018irz}. We review it here to introduce our conventions in the simplest available setup.

We consider the ten-dimensional metric
\begin{eq}
    ds^2=L^2 \dd s^2_{\text{MS}_7}+R_X^2 \dd s^2_{X_3}\mcomma
\end{eq}
and we normalize the scalar curvatures of MS$_7$ and $X_3$ as 
\begin{eq}\label[pluralequation]{eq:normalization_curvatures}
    R^{(7)}=\frac{42}{L^2}k\mcomma \qquad  R^{(3)}=\frac{6}{R_X^2}k_X\mcomma
\end{eq}
with $k,k_X\in\{-1,0,1\}$. We denote by $\text{vol}_X$ the volume form of $X_3$ and by $\text{Vol}_X$ the volume of $X_3$ measured in the metric $\dd s^2_X$, so that 
\begin{eq}
    \int_X \text{vol}_X=\text{Vol}_X\mperiod
\end{eq}
In our conventions, flux quantization for a $k$-form $F_k$ is
\begin{eq} \label{eq:conventions-flux-number}
    \int_X F_k =(2\pi \ell_s)^{k-1} n_X\mcomma
\end{eq}
with $n_X\in\Z$.
For the MS$_7\times X_3$ solution, the three-form flux is
\begin{eq}
    H_3= f_X \text{vol}_X\mcomma
\end{eq}
and the quantization condition in \cref{eq:conventions-flux-number} reads
\begin{eq}
    n_X=\frac{\text{Vol}_X}{(2\pi \ell_s)^{2}} f_X\mperiod
\end{eq}
The equations are only compatible with
\begin{eq}
    k=-1 \qquad \text{and} \qquad k_X=1\mperiod
\end{eq}
Spacetime is therefore AdS$_7\times X_3$, where $X_3$ is an Einstein manifold with positive curvature.
The dilaton and the two radii from \cref{eq:heterotic-constant-dilaton} are
\begin{eq}
    e^{\phi_0}& = 5^{\frac{1}{4}}\left(\text{Vol}_X\right)^{\frac{1}{2}}(2\pi \ell_s)^{-1} T^{-\frac{1}{2}}n_X^{-\frac{1}{2}}\mcomma \\
    \frac{(2\pi \ell_s)^2}{L^2}& = \frac{5^{\frac{1}{2}}}{12}\text{Vol}_X n_X^{-1}\mcomma \qquad
    \frac{(2\pi \ell_s)^2}{R_X^2} = 5^{\frac{1}{2}}\text{Vol}_X n_X^{-1}\mcomma
\end{eq}
where we recast the flux contribution in terms of the magnetic flux number $n_X$.
This is the AdS$_7\times S^3$ of~\cite{Mourad:2016xbk,Basile:2018irz}.
Analogous solutions are obtained by splitting AdS$_7$ as AdS$_p\times M_{7-p}$, where $M$ is an Einstein manifold with the appropriate negative curvature.

\subsubsection{\texorpdfstring{AdS$_4\times X_3 \times Y_3$}{AdS4 x X3 x Y3}}
\label{ssec:ads4X3Y3}

The first new class of vacua that we explore is a generalization of AdS$_7\times X_3$. Instead of a single internal space threaded by flux, the internal sector consists of two Einstein spaces with two different fluxes. The ten-dimensional spacetime splits into MS$_4\times X_3\times Y_3$, with a three-form flux 
\begin{eq}
    H_3= f_X \text{vol}_X+ f_Y \text{vol}_Y\mperiod
\end{eq}
A feature of this class is the selection of a four-dimensional external spacetime. Normalizing curvatures as in \cref{eq:normalization_curvatures}, the equations of motion from \cref{eq:heterotic-constant-dilaton} become
\begin{eq}  \label[pluralequation]{eq:het-XY}
    & 5 T e^{2\phi_0}=\frac{f_X^2}{R_X^6}+\frac{f_Y^2}{R_Y^6}\mcomma \qquad &&  \frac{3 k}{L^2} = -\frac{1}{10}\left(\frac{f_X^2}{R_X^6}+\frac{f_Y^2}{R_Y^6}\right)\mcomma \\
    & \frac{2 k_X}{R_X^2}=\frac{2}{5}\frac{f_X^2}{R_X^6}-\frac{1}{10}\frac{f_Y^2}{R_Y^6}\mcomma \qquad  && \frac{2 k_Y}{R_Y^2}=\frac{2}{5}\frac{f_Y^2}{R_Y^6}-\frac{1}{10}\frac{f_X^2}{R_X^6}\mperiod
\end{eq}
These are only compatible with $k=-1$, and the last two are equivalent to
\begin{eq}\label[pluralequation]{eq:het-XY-kxky}
    \frac{k_X}{R_X^2}+\frac{k_Y}{R_Y^2}=\frac{9}{2 L^2}\mcomma \qquad \frac{k_X}{R_X^2}-\frac{k_Y}{R_Y^2}=\frac{1}{4}\left(\frac{f_X^2}{R_X^6}-\frac{f_Y^2}{R_Y^6}\right)\mperiod
\end{eq}
The first of \cref{eq:het-XY-kxky} shows that $k_X$ and $k_Y$ cannot both be negative. Three cases remain.

The first case corresponds to\footnote{This case already appears in~\cite{Basile:2020xwi}.}  
\begin{eq}
    k_X=1 \qquad \text{and} \qquad k_Y=1\mcomma
\end{eq}
and thus the first of \cref{eq:het-XY-kxky} is equivalent to setting
\begin{eq}
    \frac{1}{R_X}=\frac{3}{\sqrt{2}L}\cos\theta \qquad \text{and} \qquad \frac{1}{R_Y}=\frac{3}{\sqrt{2}L}\sin\theta \mcomma
\end{eq}
in terms of an angle $\theta$. 
$\phi_0$ and $L$ are determined by the first two of \cref{eq:het-XY}. The remaining equation, the second of \cref{eq:het-XY-kxky}, is equivalent to
\begin{eq} \label{eq:het-XY++theta}
    \left(\frac{f_X}{f_Y}\right)^2= \frac{5-3\left(\sin^2\theta-\cos^2\theta\right)}{5+3\left(\sin^2\theta-\cos^2\theta\right)} \tan^6\theta\mcomma
\end{eq}
in which $f_{X,Y}$ can be replaced with the quantized fluxes $n_{X,Y}\propto f_{X,Y}$. For any choice of fluxes, there is an angle $\theta\in(0,\frac{\pi}{2})$ that solves \cref{eq:het-XY++theta}, thus providing a complete solution with $X$ and $Y$ of positive curvature. The special case with internal spheres and equal fluxes, $f_X=f_Y$, will be relevant in \cref{ssec:ads-examples}.

The second case corresponds to
\begin{eq}
    k_X=1 \qquad \text{and}\qquad  k_Y=-1\mperiod
\end{eq}
Letting
\begin{eq}
    \frac{1}{R_X}=\frac{3}{\sqrt{2}L}\cosh\theta \qquad \text{and} \qquad \frac{1}{R_Y}=\frac{3}{\sqrt{2}L}\sinh\theta\mcomma
\end{eq}
analogous manipulations link the parameter $\theta$ to the ratio of the two fluxes,
\begin{eq} \label{eq:het-XY+-theta}
    \left(\frac{f_X}{f_Y}\right)^2= \frac{5+3\left(\sinh^2\theta+\cosh^2\theta\right)}{5-3\left(\sinh^2\theta+\cosh^2\theta\right)} \tanh^6\theta\mperiod
\end{eq}
For any choice of quantized fluxes $n_{X,Y}\propto f_{X,Y}$, there is a value of $\theta\in(0,\frac{\log 3}{2})$ that solves \cref{eq:het-XY+-theta}, thus providing a solution with a four-dimensional AdS spacetime and two internal spaces with curvatures of opposite signs.

The third case corresponds to
\begin{eq}
    k_X=1 \qquad \text{and}\qquad  k_Y=0\mcomma
\end{eq}
and takes the explicit form
\begin{eq}
    R_X^2=\frac{3^{\frac{1}{2}}}{4}f_X\mcomma\qquad R_Y^2=\frac{3^{\frac{1}{2}}}{2^{\frac{4}{3}}}\left(f_X f_Y^2\right)^{\frac{1}{3}}\mcomma \qquad L^2=\frac{3^{\frac{5}{2}}}{8}f_X\mcomma \qquad e^{\phi_0}=\frac{4}{3^{\frac{3}{4}}} T^{-\frac{1}{2}}f_X^{-\frac{1}{2}}\mperiod
\end{eq}

To show that these solutions are reliable---small string coupling and large radii---it suffices to show that the AdS radius $L$ grows parametrically with the flux numbers. For generic values of $\theta$, this is an immediate consequence of the second of \cref{eq:het-XY}. For the special limits of $\theta$, $\theta\sim0$ in the first case and $\theta\sim0$ or $\theta\sim\frac{\log 3}{2}$ in the second case, reliability follows from expanding \cref{eq:het-XY++theta,eq:het-XY+-theta}, respectively.

A similar AdS$_3\times X_3 \times Y_4$ solution exists by turning on the dual seven-form flux, with $H_3=f_X \text{vol}_X$ and $H_7= f_{XY}\text{vol}_X\wedge\text{vol}_Y$.
It is only compatible with $k_X=1$ and $k_Y=-1$. We refrain from displaying it explicitly because the expressions would be more complex and yet would add little more to the overall picture.

\subsubsection{\texorpdfstring{AdS$_5\times T_{p,q}$}{AdS5 x Tpq}}
\label{ssec:Tpq}

The Freund--Rubin vacua of~\cite{Mourad:2016xbk} and the generalized version of \cref{ssec:ads4X3Y3} balance the contribution of the tadpole potential with those of the curvatures and fluxes. These are not the only options that follow this strategy. More complex internal manifolds can be employed, provided they admit a harmonic three-form. For instance, one can consider fibrations in which the harmonic three-form is not entirely parallel to the base. Here, we consider one such case, which is a generalization of $T_{p,q}$ manifolds.

Take an $S^1$ fibration over two Riemann surfaces $X$ and $Y$ with the metric
\begin{eq}
    \dd s^2=L^2 \dd s_{\text{MS}_5}+ R^2 (\dd\psi+A)^2+ R_X^2 \dd s_X^2+ R_Y^2 \dd s_Y^2 \mperiod
\end{eq}
The connection $A$ controls the $S^1$ fibration. We choose $R$ so that the coordinate $\psi$ along the circle has period $4\pi$ to make the intermediate expressions cleaner. Similar to the previous examples, we denote the curvatures of the two surfaces by $k_{X,Y}\in\{-1,0,1\}$, and we also introduce the Euler characteristics $\chi_{X,Y}$ of $X$ and $Y$.
The curvature $F=\dd A$ takes the general form $F=F_{12}e^1\wedge e^2+ F_{34}e^3\wedge e^4$ in terms of the vielbein,
\begin{eq}
    e^{1,2}=R_X \Tilde{e}^{1,2}\mcomma \qquad e^{3,4}=R_Y \Tilde{e}^{3,4}\mcomma \qquad e^{5}=R(d\psi+A)\mcomma
\end{eq}
where $\Tilde{e}$ denotes the vielbeins of the two surfaces with metrics $\dd s_{X,Y}^2$.
The quantization\footnote{The period of $\psi$ is $4\pi$; therefore, the quantization condition has an additional factor of $2$ when compared to the usual one. We use the same conventions as those of~\cite{Candelas:1989js} for the $T_{p,1}$ case.} of the curvature, which computes the first Chern class of the U(1) bundle, leads to 
\begin{eq}\label{eq:Tpq-F}
    F=\frac{2k_X p}{\chi_X R_X^2} e^1\wedge e^2+ \frac{2 k_Y q}{\chi_Y R_Y^2} e^3\wedge e^4\mcomma
\end{eq}
with $p,q\in\Z$. \Cref{eq:Tpq-F} is valid provided that $\chi\neq0$ for the two Riemann surfaces. When $\chi=0$, one can replace $\frac{\chi}{k}$ with an arbitrary positive value proportional to the volume of the torus.

The harmonic three-form on this $S^1$ fibration is 
\begin{eq}
    H_3=h\left[\frac{2k_X p}{R_X^2\chi_X} e^5\wedge e^1\wedge e^2-\frac{2k_Y q}{R_Y^2 \chi_Y} e^5\wedge e^3\wedge e^4\right]\mcomma
\end{eq}
and flux quantization from \cref{eq:conventions-flux-number} demands
\begin{eq}
    \frac{4 h R}{\ell_s^2}=n_H\in\Z\mperiod
\end{eq}
From \cref{eq:heterotic-constant-dilaton}, one finds an AdS$_5$ vacuum with radius $L$ such that
\begin{eq}\label[pluralequation]{eq:Tpq-full-equations}
    & L^{-2}=\frac{T}{8}e^{2\phi_0}\mcomma \qquad R^2=\frac{4}{5}h^2 \mcomma \qquad h^2\left(\frac{4 p^2}{\chi_X^2 R_X^4}+\frac{4 q^2}{\chi_Y^2 R_Y^4}\right)=5T e^{2\phi_0}\mcomma \\
    & \frac{k_X}{R_X^2}+\frac{k_Y}{R_Y^2}=\frac{7}{2}T e^{2\phi_0}\mcomma \qquad \frac{k_X}{R_X^2}-\frac{k_Y}{R_Y^2}=\frac{9}{10}h^2\left(\frac{4 p^2}{\chi_X^2 R_X^4}-\frac{4 q^2}{\chi_Y^2 R_Y^4}\right)\mperiod
\end{eq}
The first equation of the second line implies that $k_X $ and $k_Y$ cannot both be negative. Three separate cases remain.

The first case is when
\begin{eq}
    k_X=1\qquad \text{and}\qquad k_Y=1\mperiod
\end{eq}
One can parameterize $R_{X,Y}$ with an angle $\theta\in(0,\frac{\pi}{2})$, letting\footnote{$\chi_{X,Y}=2$ in this case, but we leave them implicit for ease of comparison with later expressions.}
\begin{eq}
    \frac{1}{R_X}=\sqrt{\frac{7T}{2}}e^{\phi_0}\cos\theta\mcomma \qquad \frac{1}{R_Y}=\sqrt{\frac{7T}{2}}e^{\phi_0}\sin\theta\mperiod
\end{eq}
The two remaining equations fix $\phi_0$ and the angle $\theta$ in terms of $p, q$, the flux $n_H$, and the two Euler characteristics:
\begin{eq}\label[pluralequation]{eq:Tpq-sphere-theta}
        e^{2\phi_0}&=\frac{8\sqrt{5}}{7^2}\frac{1}{T \ell_s^2}\left(\frac{p^2}{\chi_X^2}\cos^4\theta+\frac{q^2}{\chi_Y^2}\sin^4\theta\right)^{-1} n_H^{-1}\mcomma \\
    \left(\frac{p}{q}\right)^2& = \left(\frac{\chi_X}{\chi_Y}\right)^2\frac{9+7\cos(2\theta)}{9-7\cos(2\theta)}\tan^4\theta\mperiod
\end{eq}
For any $p,q\neq0$ there exists a $\theta$ that solves the second of \cref{eq:Tpq-sphere-theta}. Hence, these vacua with an internal $T_{p,q}$ space always exist, and the complete solution in terms of the free (integer) parameters $p,q$, and $n_H$ reads
\begin{eq}
    &\frac{R}{\ell_s}=\frac{n_H^{\frac{1}{2}}}{(20)^{\frac{1}{4}}}\mcomma\qquad  \frac{h}{\ell_s}=\frac{(20)^{\frac{1}{4}}n_H^{\frac{1}{2}}}{4}\mcomma \qquad \frac{L}{\ell_s}=\frac{7}{2\cdot 5^{\frac{1}{4}}}\left(p^2\cos^4\theta+q^2\sin^4\theta\right)^{\frac{1}{2}}n_H^{\frac{1}{2}}\mcomma \\
    & e^{\phi_0}=\frac{2^{\frac{5}{2}}\cdot 5^{\frac{1}{4}}}{7}(T \ell_s^2)^{-\frac{1}{2}}\left(p^2\cos^4\theta+q^2\sin^4\theta\right)^{-\frac{1}{2}}n_H^{-\frac{1}{2}}\mcomma \\
    &\frac{R_X}{\ell_s}=\frac{7^{\frac{1}{2}}}{4\cdot 5^{\frac{1}{4}}}\left(p^2\cos^4\theta+q^2\sin^4\theta\right)^{\frac{1}{2}}n_H^{\frac{1}{2}}(\cos\theta)^{-1}\mcomma \\
    &\frac{R_Y}{\ell_s}=\frac{7^{\frac{1}{2}}}{4\cdot 5^{\frac{1}{4}}}\left(p^2\cos^4\theta+q^2\sin^4\theta\right)^{\frac{1}{2}}n_H^{\frac{1}{2}}(\sin\theta)^{-1}\mcomma
\end{eq}
where $\theta$ is determined by
\begin{eq}
    \frac{9+7\cos(2\theta)}{9-7\cos(2\theta)}\tan^4\theta=\left(\frac{p}{q}\right)^2 \mperiod
\end{eq}

The second case is when
\begin{eq}
        k_X=1\qquad \text{and}\qquad k_Y=-1\mperiod 
\end{eq}
Here, one can parameterize $R_{X,Y}$ as
\begin{eq}
    \frac{1}{R_X}=\sqrt{\frac{7T}{2}}e^{\phi_0}\cosh\theta\mcomma \qquad \frac{1}{R_Y}=\sqrt{\frac{7T}{2}}e^{\phi_0}\sinh\theta\mperiod
\end{eq}
The two remaining equations fix $\phi_0$ and $\theta$ in terms of $p,q$, and the flux number. The complete solution is an $S^1$ fibration over a two-sphere and a Riemann surface of genus $g>0$. It reads
\begin{eq}
    &\frac{R}{\ell_s}=\frac{n_H^{\frac{1}{2}}}{(20)^{\frac{1}{4}}}\mcomma\qquad  \frac{h}{\ell_s}=\frac{(20)^{\frac{1}{4}}n_H^{\frac{1}{2}}}{4}\mcomma \qquad \frac{L}{\ell_s}=\frac{7}{2\cdot 5^{\frac{1}{4}}}\left(p^2\cosh^4\theta+\frac{4}{\chi_Y^2}q^2\sinh^4\theta\right)^{\frac{1}{2}}n_H^{\frac{1}{2}}\mcomma \\
    & e^{\phi_0}=\frac{2^{\frac{5}{2}}\cdot 5^{\frac{1}{4}}}{7}(T \ell_s^2)^{-\frac{1}{2}}\left(p^2\cosh^4\theta+\frac{4}{\chi_Y^2}q^2\sinh^4\theta\right)^{-\frac{1}{2}}n_H^{-\frac{1}{2}}\mcomma \\
    &\frac{R_X}{\ell_s}=\frac{7^{\frac{1}{2}}}{4\cdot 5^{\frac{1}{4}}}\left(p^2\cosh^4\theta+\frac{4}{\chi_Y^2}q^2\sinh^4\theta\right)^{\frac{1}{2}}n_H^{\frac{1}{2}}(\cosh\theta)^{-1}\mcomma \\
    &\frac{R_Y}{\ell_s}=\frac{7^{\frac{1}{2}}}{4\cdot 5^{\frac{1}{4}}}\left(p^2\cosh^4\theta+\frac{4}{\chi_Y^2}q^2\sinh^4\theta\right)^{\frac{1}{2}}n_H^{\frac{1}{2}}(\sinh\theta)^{-1}\mcomma
\end{eq}
    where $\theta$ is determined by
\begin{eq}
    \frac{9+7\cosh(2\theta)}{9-7\cosh(2\theta)}\tanh^4\theta=\frac{\chi_Y^2}{4}\left(\frac{p}{q}\right)^2 \mperiod
\end{eq}
This has a solution for any choice of non-zero $p$ and $q$, with $\theta\in(0,\frac{1}{2}\arccosh{\frac{9}{7}})$.

The third case is
\begin{eq}
    k_X=1\qquad \text{and}\qquad k_Y=0\mperiod 
\end{eq}
One can still use \cref{eq:Tpq-full-equations} by normalizing the volume of the torus to $1$. Formally, one replaces $2\pi\chi_Y\to k_Y$ and then takes the $k_Y\to 0$ limit. The solution is
\begin{eq}
    & \frac{R}{\ell_s}\sim\frac{h}{\ell_s}\sim n_H^{\frac{1}{2}}\mcomma \qquad \frac{L}{\ell_s}\sim\frac{R_X}{\ell_s}\sim  n_{H}^{\frac{1}{2}} p \mcomma \qquad e^{\phi_0}\sim \frac{n_H^{-\frac{1}{2}}}{p\sqrt{T\ell_s^2}}\mcomma \qquad \frac{R_Y}{\ell_s}\sim  n_H^{\frac{1}{2}}\sqrt{pq}\mperiod 
\end{eq}
We omit numerical factors because this case can be obtained from the $T_{p,q}$ vacua after expanding the expressions as $\theta\to 0$, rescaling $R_Y$ by a factor $\sqrt{4\pi}$, and taking into account that the relation between $p,q$, and $\theta$ becomes $(p/q)^2\sim 8 \theta^4$.

The three cases are thus parameterized by three integers, $p,q$, and $n_H$, and by the genera of the Riemann surfaces. 
The AdS spacetime is five-dimensional. 

One could try to extend the setup in this section by including an additional $S^1$ fibration, but we will not explore this possibility in this paper.

Looking more broadly at the problem,
the main obstacle to obtaining new solutions is the necessity of having a harmonic 3-form.
Inspired by $T_{p,q}$, one could consider more general homogeneous spaces as internal manifolds. However, a classification along the lines of~\cite{Chapline:1982wy,Castellani:1983yg,lauret-will} leaves only a few possibilities for an internal space with $H^3=\Z$ of sufficiently low dimension. Other examples of the type $G/H$, with $G$ semisimple and compact, are SU(2) for $n=3$, and SU(3) for $n=8$. This leaves no ingredients to build a four-dimensional external space, except for the product of two three-dimensional manifolds that we have already explored. Motivated by this, in \cref{sec:cpt-H-F} we include gauge fields, opening new possibilities and obtaining four-dimensional vacua with tadpole potentials.

Before delving into this aspect, in \cref{ssec:half-loop}, we comment on the analogous solutions of the ten-dimensional non-supersymmetric orientifolds.

\subsection{Half-loop orientifold}\label{ssec:half-loop}

Similar solutions are possible for the two non-supersymmetric orientifolds, Sugimoto's USp$(32)$ model, and the type 0'B string. All fluxes come from Ramond--Ramond fields, and the tadpole potential has a half-loop origin, so that
\begin{eq}
    \beta_k=0\mcomma \qquad \gamma=-1 \mperiod
\end{eq}
The allowed values of $k$ are $k=3$ for the Sugimoto model and $k=1$, $3$, and $5$ for the type 0'B theory. In this section, we replace some of the fluxes with their duals, so that they are always internal.

\Cref{eq:constant-dilaton} are equivalent to
\begin{eq}\label[pluralequation]{eq:constant-dilaton-orientifold}
    T e^{-\phi_0} & = \sum_k \frac{k-5}{6} F_k^2 \mcomma \\
    e^{-2\phi_0} R_{MN} & = \sum_k \frac{1}{2}\iota_M F_k \cdot \iota_N F_k-\frac{k+1}{24}F_k^2 g_{MN} \mperiod
\end{eq}
We choose to focus on $F_3$ and its dual $F_7$ because these are the universal contributions that both orientifolds have. At the end of this section, we will comment on further options in type 0'B when one includes the self-dual $F_5$.

The positive sign of $T$ requires the presence of some $F_7$ magnetic flux.
The maximum value for the dimension of the external spacetime is therefore $D=3$.
The simplest possibility is to take a three-dimensional maximally symmetric space, MS$_3$, and a single internal seven-dimensional Einstein manifold. As in the heterotic case, this solution has already been found in~\cite{Gubser:2001zr,Mourad:2016xbk}. It consists of the ten-dimensional metric
\begin{eq}
    \dd s^2=L^2 \dd s_{\text{AdS}_3}^2 + R_X^2 \dd s_{X_7}^2\mcomma
\end{eq}
with $F_7$ flux
\begin{eq}
    F_7=f_X \text{vol}_X\mcomma
\end{eq}
and with a seven-dimensional internal space of positive curvature, $k_X=1$.
In terms of the flux number, the complete solution is
\begin{eq}
    e^{\phi_0}&=2^{\frac{7}{4}}\cdot 3 \left(4\pi^2\ell_s^2 T\right)^{-\frac{3}{4}}\left(\text{Vol}_X\right)^{\frac{1}{4}} n_X^{-\frac{1}{4}}\mcomma \\
    \frac{4\pi^2 \ell_s^2}{L^2}&=2^{\frac{3}{4}}\cdot 3 \left(4\pi^2\ell_s^2 T\right)^{\frac{1}{4}}\left(\text{Vol}_X\right)^{\frac{1}{4}} n_X^{-\frac{1}{4}}\mcomma \\
    \frac{4\pi^2 \ell_s^2}{R_X^2}&=2^{-\frac{1}{4}} \left(4\pi^2\ell_s^2 T\right)^{\frac{1}{4}}\left(\text{Vol}_X\right)^{\frac{1}{4}} n_X^{-\frac{1}{4}}\mperiod
\end{eq}

\subsubsection{\texorpdfstring{AdS$_3\times X_3 \times Y_4$}{AdS3 x X3 x Y4}}
\label{ssec:ads3X3Y4}

A generalization, consistent with the first of \cref{eq:constant-dilaton-orientifold}, is to consider the product MS$_3\times X_3\times Y_4$, with metric and fluxes
\begin{eq}
    \dd s^2& = L^2 \dd s_{\text{MS}_3}^2+ R_X^2 \dd s_{X_3}^2 + R_Y^2 \dd s_{Y_4}^2\mcomma \\
    F_3&=f_X\text{vol}_X\mcomma \\ F_7&=f_{XY}\text{vol}_X\wedge \text{vol}_Y\mperiod
\end{eq}
In this case, the equations are only compatible with
\begin{eq}
    k=-1\mcomma \qquad k_X=1\mcomma \qquad k_Y=1\mcomma
\end{eq}
and lead to the curvature radii
\begin{eq}\label[pluralequation]{eq:orientifold-solution-ads-s-s}
    \frac{1}{L^2}&=\frac{T}{4}e^{\phi_0}\frac{2 f_{XY}^2\left(\frac{T}{6}e^{\phi_0}\right)^4+f_X^2}{f_{XY}^2\left(\frac{T}{6}e^{\phi_0}\right)^4-f_X^2}\mcomma \\
    \frac{1}{R_X^2}&=\left(\frac{3 T e^{-\phi_0}}{f_{XY}^2\left(\frac{T}{6}e^{\phi_0}\right)^4-f_X^2}\right)^{\frac{1}{3}}\mcomma \\
    \frac{1}{R_Y^2}&=\frac{T}{6}e^{\phi_0}\mperiod
\end{eq}
The dilaton $\phi_0$ is obtained by solving 
\begin{eq}\label{eq:orientifold_ads3_dilaton}
    \frac{\left[f_{XY}^2\left(\frac{T}{6}e^{\phi_0}\right)^4+2f_X^2\right]^3}{\left[f_{XY}^2\left(\frac{T}{6}e^{\phi_0}\right)^4-f_X^2\right]^2} \frac{T^2}{192} e^{4\phi_0}=1\mperiod
\end{eq}
One can then replace $f_{XY}$ and $f_X$ with the flux numbers, given by
\begin{eq}
    n_{XY}=f_{XY}\frac{\text{Vol}_X\text{Vol}_Y}{(2\pi\ell_s)^6}\mcomma \qquad n_X=f_X\frac{\text{Vol}_X}{(2\pi\ell_s)^2}\mperiod
\end{eq}
Note that the denominators in \cref{eq:orientifold-solution-ads-s-s} must be positive, and thus the flux numbers are constrained by the sign of $T$. Consequently, determining the range of $n_{X,XY}$ in which the solution is reliable is more delicate than in the previous examples. One option is to work in a regime where these denominators are small, which lead to an approximate solution to \cref{eq:orientifold_ads3_dilaton},
\begin{eq}
    \left(\frac{T}{6}e^{\phi_0}\right)^4\sim \frac{f_X^2}{f_{XY}^2}\left(1+\frac{27}{2T}\frac{f_X^2}{f_{XY}}\right)\mperiod
\end{eq}
The vacuum is then reliable in the window $n_X^2\ll n_{XY}\ll n_X^3$.

\subsubsection{\texorpdfstring{AdS$_3\times X_2 \times Y_5$}{AdS3 x X2 x Y5}}
\label{ssec:ads3X2Y5}

Type 0'B string theory has further options available by turning on $F_5$ and $F_1$ fluxes.
The former is the simplest because the first of \cref{eq:constant-dilaton-orientifold} does not change.
The most basic type of vacuum with $F_5$ and $F_3$ is of the form MS$_3\times X_2\times Y_5$, with
\begin{eq}
    F_7&= f_{XY}\text{vol}_X\wedge \text{vol}_Y\mcomma \\
    F_5&= f_Y\left[\text{vol}_Y+ \frac{L^3 R_X^2}{R_Y^5}\text{vol}_{\text{MS$_3$}}\wedge \text{vol}_X\right]\mperiod
\end{eq}
The five-form must be self-dual, and therefore one must take into account the presence of additional factors of $2$ in the equations of motion. These are found to be consistent with
\begin{eq}
    k=-1\mcomma\qquad k_Y=1\mcomma
\end{eq}
and the sign of the curvature of $X_2$ depends on the two fluxes.
The complete solution can be obtained from
\begin{eq}
    & 3 T e^{-\phi_0}=\frac{f_{XY}^2}{R_X^4 R_Y^{10}}\mcomma \qquad &&\frac{1}{L^2} =\frac{T}{2}e^{\phi_0}\left[1+\frac{3}{4}\frac{f_Y^2}{f_{XY}^2}R_X^4\right]\mcomma \nonumber \\
    & \frac{k_X}{R_X^2}=\frac{T}{2}e^{\phi_0}\left[1-\frac{3}{2}\frac{f_Y^2}{f_{XY}^2}R_X^4\right]\mcomma \qquad &&\frac{1}{R_Y^2}=\frac{T}{8}e^{\phi_0}\left[1+\frac{3}{2}\frac{f_Y^2}{f_{XY}^2}R_X^4\right]\mcomma
\end{eq}
where the third equation determines $R_X$. An option is to take $\frac{f_Y^2}{f_{XY}^2}R_X^4\ll 1$, which leads to a reliable vacuum with $k_X=1$ when $n_{XY}^3\gg n_Y^4$.

\section{\texorpdfstring{Kaluza--Klein approach to the ${\rm SO}(16)\times {\rm SO}(16)$ heterotic string}{Kaluza--Klein approach to the so(16)xso(16) heterotic string}}
\label{sec:cpt-H-F}

From now on, we focus on the heterotic ${\rm SO}(16)\times {\rm SO}(16)$ string. We have already found an AdS$_4$ vacuum for it in \cref{ssec:ads4X3Y3}, but we now expand our scope and add a background for the gauge field. We will comment below on the issues that arise when trying to extend the results of this section to the orientifold theories.

To simplify the equations, we introduce the notation 
\begin{equation}\label{eq:trTr}
    {\rm tr}F^2\equiv -\frac{\ell_s^2}{60}{\rm Tr}F^2 \,,
\end{equation}
where ${\rm Tr}$ denotes the trace in the adjoint. We include the minus sign in \cref{eq:trTr} because we use the mathematical convention for the gauge algebra generators, in which the structure constants are real. This causes the Killing form ${\rm Tr}(T_I T_J)\equiv-k_{16}\delta_{IJ}$ to be negative definite, and would make wrong-looking signs appear in the action and equations of motion. In this language, the two-derivative action is
\begin{eq}
    S=\frac{2\pi}{(2\pi \ell_s)^8}\int \sqrt{-g}\left[e^{-2\phi}\left(R+4(\d\phi)^2-\frac12 H^2 - \frac12 {\rm tr} F^2 \right) - T \right] \mperiod
\end{eq}
The equations of motion and Bianchi identities read
\begin{eq}\label[pluralequation]{eq:het-eom-gauge}
    &R_{MN}  + 2 \nabla_M \nabla_N \phi + \frac{T}{2} e^{2\phi} g_{MN}- \frac{1}{2}\iota_M H \cdot \iota_N H - \frac{1}{2} {\rm tr}\, \iota_M F \cdot \iota_N F =0 \mcomma \\
    & R + 4\nabla^2 \phi - 4(\d\phi)^2 - \frac{1}{2} H^2 -\frac{1}{2} \text{tr} F^2 = 0 \mcomma \\
    & \dd H  = -\frac{1}{2} \text{tr} F\wedge F\mcomma \qquad \dd(e^{-2\phi}\hodge H)=0\mcomma \\
    & \dd_A F  = 0 \mcomma \qquad e^{2\phi}\dd_A\left(e^{-2\phi}\hodge F\right)+F\wedge \hodge H = 0\mcomma
\end{eq}
where 
\begin{eq}
    H=\dd B -\frac{1}{2}\omega_{YM}\mperiod
\end{eq}
The heterotic equations are often given with Riemann-squared terms, especially in the three-form Bianchi identity. In fact, there is an infinite series of higher-derivative corrections; the Riemann-squared terms are singled out when $F$ is taken to be small, so that its contribution can cancel with them. This leads to the \emph{standard embedding}. In contrast, here we take $F$ to be large, and the higher-derivative corrections are negligible, as is more often the case in flux compactifications.

We consider direct products ${\rm AdS}_{10-d} \times M_d$, with a direct (unwarped) product metric 
\begin{equation}
    \dd s^2 = \dd s^2_{{\rm AdS}_{10-d}}+ \dd s^2_d\,.
\end{equation}
As in the previous section, we take the dilaton to be constant. The form fields are purely internal. In this situation, the last two lines in \cref{eq:het-eom-gauge} remain formally identical, with the understanding that the forms and the Hodge star are purely internal, and $e^{\phi_0}=g_s$. The first two in \cref{eq:het-eom-gauge} become
\begin{subequations}    \label[pluralequation]{eq:heterotic_system}
\begin{align}
    \label[pluralequation]{eq:TL}
    R_{\mu\nu}&=-\frac{T g_s^2}{2}  g_{\mu\nu}\mcomma
    \qquad\qquad
    T g_s^2 = \frac{1}{5}H^2+\frac{1}{10} {\rm tr} F^2 \mcomma
    \\
    \label{eq:intR}
        R_{mn} &= \frac12 \iota_m H \cdot\iota_n H -\frac1{10} H^2 g_{mn} +\frac12 {\rm tr}\,\iota_m F \cdot\iota_n F -\frac1{20} {\rm tr} F^2 g_{mn}
\mperiod
\end{align}
\end{subequations}
\Cref{eq:TL} fix the values of $g_s$ and of the cosmological constant $\Lambda$. The latter is negative, thus explaining our previously declared focus on AdS. Note that for the orientifold theories, the right-hand side of the second of \cref{eq:TL} is negative if the form fields are internal (see \cref{eq:constant-dilaton}). This would require an AdS space of dimension at most three, with at least an external $H$-flux. We will not explore this possibility in this work.

Our strategy to solve the system in \cref{eq:heterotic_system} is to generate a gauge field with a gauge group $H<{\rm SO}(16)\times {\rm SO}(16)$ by a formal dimensional reduction on a $H$-principal fibration. Similar ideas have been used in the literature before, both in physics and in mathematics~\cite{Duff:1985cm,Lechtenfeld:2010dr,garcia2024pluriclosed}. We explain the idea in general and schematic terms in \cref{sub:red}; in \cref{sec:hom}, we spell it out in detail when the internal space is $M_d= G/H$, which is the case where we found explicit solutions.

\subsection{Reductions and Bianchi identities}
\label{sub:red}

Consider a fiber bundle $F \hookrightarrow E\to M$ with fiber $F$, with a metric that is invariant under Killing vectors $K_\alpha$, with commutators $[K_\alpha,K_\beta]=-f^\gamma{}_{\alpha \beta}K_\gamma$. A natural class of metrics on the total space $E$ of the bundle is 
\begin{equation}\label{eq:bun-met}
	\dd s^2_E \equiv g^B_{mn}\dd x^m \dd x^n +  g^F_{uv} D y^u D y^v\,,
\end{equation}
with $D y^u\equiv \dd y^u + K^u_\alpha A^\alpha$. The $A^\alpha$ are connection forms on the base $M$, with field strength $F=\dd A + A \wedge A= \dd A + \frac12 [A,A]$, $F^\alpha=\dd A^\alpha + \frac12 f^\alpha{}_{\beta \gamma} A^\beta \wedge A^\gamma$. In this situation, the Ricci scalar reduces as
\begin{equation}\label{eq:EH-red}
    R_E = R_B + R_F -\frac{1}{2} \, K_\alpha \cdot K_\beta \, F^\alpha \cdot F^\beta\,,
\end{equation}
where the dots denote the usual pointwise inner products: $F^\alpha \cdot F^\beta= \frac12 F^\alpha_{ij}F^{\beta ij}$, while $K_\alpha\cdot K_\beta= K_\alpha^u K_{\beta u}$. 

In \cref{eq:EH-red} we see one of the most remarkable aspects of dimensional reductions: the Einstein--Hilbert term generates an extra gauge field. The Yang--Mills equations of motion, $\dd *F+ [A,*F]=0$, appear as the Einstein equations $R^E_{iu}=0$. However, the Einstein equations $R^E_{ij}= 0$ are $R^M_{ij} -\frac12 \, K_\alpha\cdot K_\beta \ \iota_i F^\alpha\cdot \iota_j F^\beta=0$. In general $K_\alpha\cdot K_\beta$ are not constant on $F$, so these equations do not really reduce to $M$. 

One possibility to avoid this issue is to take $F=H$, a group; in other words, to take the bundle to be principal. Taking the Killing vectors to be the generators $r_\alpha$ of the right multiplication action on $H$, the index $\alpha$ has the same range as $u$---namely $1,\ldots,{\rm dim}H$---and $g^F_{uv} D y^u D y^v= {\rm tr}\lambda^2$, where
\begin{equation}\label{eq:l}
   \lambda=  h^{-1} \dd h + h^{-1}A h\,.
\end{equation}
This satisfies 
\begin{equation}\label{eq:dl-F}
    \dd \lambda + \lambda^2 = h^{-1} F h\,.
\end{equation}

The heterotic theory also has a three-form $H$. A natural Ansatz similar to the metric in \cref{eq:bun-met} involves a Chern--Simons term:
\begin{equation}\label{eq:bun-H}
    H_E= H_B +\frac12 {\rm tr} \left(\lambda \wedge \dd \lambda +\frac23 \lambda^3\right)\,.
\end{equation}
By \cref{eq:dl-F}, the Bianchi identity $\dd H_E=0$ now reduces to $\dd H_B=-\frac12 {\rm tr} F\wedge F$, as in \cref{eq:het-eom-gauge}. More generally, it was shown in \cite{Duff:1985cm} that with \cref{eq:bun-H,eq:bun-met}, the equations of motion of the bosonic string in $d=506$ dimensions with $H=E_8\times E_8$ or ${\rm SO}(32)$ reduce to the bosonic equations of heterotic supergravity with gauge group $H$. 

Our equations~\eqref{eq:het-eom-gauge} are slightly different from those of the ordinary heterotic supergravity because of the one-loop tadpole $T$, as well as the different gauge group. However, we will show below that a similar strategy as in \cite{Duff:1985cm} works in our case, and we will use it to find AdS vacua. Indeed, \cref{eq:TL} show that with $T=0$ such vacua would not exist.

Since the Bianchi identity for $H$ on the base $M$ becomes the closure of a three-form on $E$, it is natural to look for spaces with simple metrics that have a natural harmonic three-form. This suggests taking $E=G$, another group. The bundle becomes $H \hookrightarrow G \to G/H$ and the internal space for our AdS vacua is $M= G/H$. Of course, $H$ is now a subgroup of ${\rm SO}(16)\times {\rm SO}(16)$.

In the next sections, we will show in detail how the strategy works. 

\subsection{Homogeneous spaces}
\label{sec:hom}

\subsubsection{Invariant forms}
\label{sub:inv}

Let $T_I$ be the generators of $G$, with $[T_I,T_J]= f^K{}_{IJ} T_K$. The structure constants satisfy the Jacobi identities:
\begin{equation}\label{eq:jacobi}
    f^I{}_{J[K} f^J{}_{LM]}=0\,.
\end{equation}
The one-forms $g^{-1}\dd g = \lambda^I T_I = \lambda^i T_i + \lambda^\alpha T_\alpha$ are left-invariant: they do not change under the left-multiplication maps $L_{g_0}: g\mapsto g_0 g$, for any $g_0$. They satisfy
\begin{equation}\label{eq:dlI}
	\dd \lambda^I = - \frac12 f^I{}_{JK} \lambda^J \wedge \lambda^K\,.
\end{equation}
They are not invariant under right-multiplications $R_{g_0}: g \mapsto g g_0$. Their Lie derivative under the infinitesimal generators $r_I$ of such maps reads
\begin{equation}\label{eq:Lrl}
	L_{r_I} \lambda^J= - f^J{}_{IK}\lambda^K\,.
\end{equation}

The Killing quadratic form is defined by
\begin{equation}\label{eq:killing}
    \mathrm{Tr}\left(T_I T_J\right)= f^L{}_{IK} f^K{}_{JL}= k_{IJ}\mperiod
\end{equation}
From now on, we denote by Tr the trace in the adjoint of $G$.
For simple compact groups $G$, the Killing form is negative definite; by a change of basis, one may then set
\begin{equation}
    k_{IJ}=-k \delta_{IJ}
\end{equation}
for some $k>0$. 
It would be natural to use the Killing form to raise and lower indices, but we find it easier to define $\tilde f_{IJK}\equiv \delta_{IL} f^L{}_{JK}$. 

For cleaner formulas, we introduce the shorthand notation
\begin{equation}
	\lambda^{I_1\ldots I_k} \equiv \lambda^{I_1} \wedge\ldots \wedge \lambda^{I_k}\,.
\end{equation}

We now introduce a subgroup $H<G$, and the coset space $G/H$. We divide the generators into $T_\alpha$ that generate $\mathfrak{h}<\mathfrak{g}$, and $T_i$ that generate its complement $\mathfrak{k}$, which we take to be orthogonal to $\mathfrak{h}$ with respect to the Killing form; moreover, it can be chosen to satisfy $[\mathfrak{h},\mathfrak{k}]\subset \mathfrak{k}$ when $H$ is compact or semisimple. In other words, $f^\alpha{}_{i \beta}=0$. 

Specializing the indices of \cref{eq:jacobi}, one finds several consequences; the following will be useful later: 
\begin{subequations}\label[pluralequation]{eq:jacobi-red}
\begin{alignat}{3}
    \label[pluralequation]{eq:jacobi-red-i}&f^i{}_{p[j}f^p{}_{kl]}+f^i{}_{\alpha[j}f^\alpha{}_{kl]}=0\,,\quad
        &&f^l{}_{\alpha k} f^k{}_{ij}= 2 f^l{}_{k[i} f^k{}_{j]\alpha}
		\, ,\quad
		 &&f^j{}_{i \gamma} f^\gamma{}_{\alpha \beta}= 2  f^j{}_{k[\alpha} f^k{}_{\beta] i}\,,\\
           \label[pluralequation]{eq:jacobi-red-a}&f^\epsilon{}_{\delta[\alpha} f^\delta{}_{\beta \gamma]}=0 \, ,\quad
		 &&f^\alpha{}_{\beta \gamma} f^\gamma{}_{ij}= 2f^\alpha{}_{k[i} f^k{}_{j] \beta}\,,
        \quad
        &&f^\alpha{}_{l[i}f^l{}_{jk]}=0\,.
\end{alignat}
\end{subequations}
Moreover, in terms of $\mathfrak{k}$ and $\mathfrak{h}$, (\ref{eq:dlI}) reduces to
\begin{subequations}\label[pluralequation]{eq:dlia}
\begin{align}
    \label{eq:dli}
	\dd \lambda^i &= -\frac12 f^i{}_{jk} \lambda^j \wedge \lambda^k - f^i{}_{\alpha j} \lambda^\alpha \wedge \lambda^j \,,\\
	\label{eq:dla}
	\dd \lambda^\alpha &= -\frac12 f^\alpha{}_{jk} \lambda^j \wedge \lambda^k -\frac12 f^\alpha{}_{\beta \gamma} \lambda^\beta \wedge \lambda^\gamma\,.
\end{align}
\end{subequations}

Not all of the $\lambda^I$ make sense as forms on $G/H$. Recall that a tensor with lower indices (such as a form) on a bundle is the pullback of a tensor on the base if it is horizontal and invariant, namely annihilated by $\iota_v$ and $L_v$ for $v$ a vector along the fiber. In our case, horizontality means selecting the $\lambda^i$ and their tensor products. The vertical vectors are the right-multiplication generators $r_\alpha$; by \cref{eq:Lrl}, their invariance would require $f^i{}_{\alpha j}=0$, which is rarely the case. Therefore, the $\lambda^i$ are usually not the pullback of forms on $G/H$. However, consider a higher-rank form $A_k = \frac1{k!} A_{i_1\ldots i_k} \lambda^{i_1\ldots i_k}$. This is horizontal, $\iota_{r_\alpha} A_k=0$; it is invariant if $L_{r_\alpha} A_k=\iota_{r_\alpha} \dd A_k=0$. In other words, $\dd A_k$ should have no $\lambda^\alpha$. In view of \cref{eq:dli}, this becomes the concrete condition
\begin{equation}\label{eq:a-inv}
	f^j{}_{\alpha[i_1} A_{i_2\ldots i_k]j}=0\,.
\end{equation}
This more general condition often admits solutions, as we will see in the examples below.

For more details on invariant forms on coset spaces, see \cite[Sec.~3]{Koerber:2008rx} or \cite[Sec.~4.4]{Tomasiello:2022dwe}.

\subsubsection{Three-form}
\label{sub:dh}

We now show how to solve the three-form Bianchi equation. 

It is possible to parameterize the general element $g\in G$ as $g= k h$, where $k$ and $h$ are the exponentials of elements of $\mathfrak{k}$, $\mathfrak{h}$. Then, $g^{-1}\dd g= h^{-1}\dd h + h^{-1}(k^{-1}\dd k) h$; in particular, $\lambda^i T_i = h^{-1}(k^{-1}\dd k)_{\mathfrak{k}}h$, and \cref{eq:l} becomes
\begin{equation}\label{eq:l-A}
 	\lambda^\alpha T_\alpha = h^{-1} \dd h + h^{-1}A h\,,
\end{equation} 
with $A\equiv (k^{-1}\dd k)_{\mathfrak{h}}$; this is by construction a connection on the principal bundle $H \hookrightarrow G\to G/H$.  Using \cref{eq:dl-F}, one sees that $\dd \lambda^\alpha = -\frac12 f^\alpha{}_{\beta \gamma} \lambda^\beta \wedge \lambda^\gamma + (h^{-1} F h)^\alpha$. By comparing with \cref{eq:dla}, we obtain
\begin{equation}\label{eq:hFh}
	 h^{-1} F h = -\frac12 f^\alpha{}_{jk} \lambda^{jk} T_\alpha\,.
\end{equation}
This $F$ is now a curvature on the principal bundle, so its flux quantization properties are automatically satisfied.

Using \cref{eq:l-A} also gives 
\begin{equation}\label{eq:d(hFh)}
    \dd (h^{-1} F h ) + [\lambda^\alpha T_\alpha,h^{-1} F h] = h^{-1} (\dd F + [A, F]) h\,.
\end{equation}
Moreover, using \cref{eq:jacobi,eq:dli}---in particular, the last two of \cref{eq:jacobi-red-a}---we obtain
\begin{eq}
    \dd(f^\alpha{}_{jk}\lambda^{jk}) &=
    2f^\alpha{}_{jk}\dd \lambda^j \wedge \lambda^k = -
    f^\alpha{}_{jk} (f^j{}_{lp} \lambda^{lp}+2 f^j{}_{\beta l}\lambda^\beta \wedge \lambda^l) \wedge \lambda^k \\
    &= f^\alpha{}_{\gamma\beta}f^\gamma{}_{kl} \lambda^{kl}  \wedge \lambda^\beta\,.
\end{eq}
Together with the two previous equations, this shows that the Bianchi identity, 
\begin{eq}
    \dd F +[A,F]=0\mcomma
\end{eq} 
holds automatically. Note that each $F^\alpha$ is horizontal and invariant, thus defining a form on $M$.

Now consider the three-form $\mathrm{Tr}(g^{-1}\dd g)^3= -\frac k2 \tilde f_{IJK} \lambda^{IJK}$ on $G$. It is closed: 
\begin{equation}\label{eq:dg-g3}
	\frac13\dd \mathrm{Tr}(g^{-1}\dd g)^3 = -\mathrm{Tr}(g^{-1}\dd g)^4  = -\mathrm{Tr}(T_I T_J T_K T_L) \lambda^{IJKL}=0\,,
\end{equation}
where the last step follows from the cyclicity of the trace. We now investigate how this form reduces to $G/H$. Splitting the indices and using again \cref{eq:dla}, we can write 
\begin{equation}\label{eq:g-g3}
	-\frac1{3k}\mathrm{Tr}(g^{-1}\dd g)^3 = \frac16 \tilde f_{ijk} \lambda^{ijk} -\lambda^\alpha \wedge \dd \lambda_\alpha -\frac13  f_{\alpha \beta\gamma} \lambda^{\alpha\beta\gamma}\,.
\end{equation}
(We have lowered the index $\alpha$ with $\delta_{\alpha\beta}$; we will do so from now on.)
We automatically get an expression of the form of \cref{eq:bun-H}, with its distinctive Chern--Simons-like term. The latter does not yet define a form on $G/H$, since the $\lambda^\alpha$ are not horizontal. This is related to the Chern--Simons form not being gauge invariant.

Taking the exterior differential of \cref{eq:g-g3} and recalling \cref{eq:dg-g3}, we see that
\begin{align}\label{eq:3form0}
	\frac16 \dd (\tilde f_{ijk} \lambda^{ijk} ) &= \dd \lambda_\alpha \wedge \dd \lambda^\alpha + f_{\alpha \beta \gamma} \dd \lambda^\alpha \wedge \lambda^{\beta\gamma}  \\
	&=  -\frac14 f_{\alpha\beta\gamma} f^{\alpha}{}_{\delta\epsilon} \lambda^{\beta \gamma \delta\epsilon} + (h^{-1} F h)_\alpha \wedge (h^{-1} F h)^\alpha= -\frac1{k}\mathrm{Tr}(h^{-1}F h \wedge h^{-1} F h )\nonumber\\ 
    &= -\frac1{k}
    \mathrm{Tr}(F \wedge F)= F_\alpha \wedge F^\alpha\,.\nonumber
\end{align}
In the first step, we used \cref{eq:dlia}; in the second, the Jacobi identity $f^\epsilon{}_{\delta[\alpha} f^\delta{}_{\beta \gamma]}=0$ from the first of \cref{eq:jacobi-red-a}. \Cref{eq:3form0} can also be obtained using \cref{eq:dli} and two of \cref{eq:jacobi-red}.  
We already saw that $F_\alpha$ are horizontal and invariant; \cref{eq:3form0} now shows that the same holds for $\tilde f_{ijk} \lambda^{ijk}$. 

For future reference, we also notice that the symmetric tensor
\begin{equation}
	q_{ij}\equiv f_{ikl}f_j{}^{kl} 
\end{equation} 
is invariant. Using \cref{eq:jacobi-red-i} repeatedly,
\begin{eq}\label{eq:q-inv}
	f^k{}_{\alpha i} q_{kj}&= f^k{}_{\alpha i} f_{klm} f_j{}^{lm}= -2 f_{ikl} f^k{}_{m \alpha} f_j{}^{lm} = -2 f_i{}^{kl}f_{jml} f^m{}_{k \alpha}= f_i{}^{kl}f^m{}_{\alpha j} f_{mlk} \\
    &=- f^k{}_{\alpha j} q_{ik} \,.
\end{eq}

\subsubsection{Riemann tensor}
\label{sub:riemann}

We now review a formula for the Riemann tensor of homogeneous spaces; see, for instance, \cite[Sec.~7.C]{besse} and \cite[(4.8)]{Castellani:1983tb} for equivalent expressions in different formulations.

We introduce a metric and the corresponding vielbein:
\begin{equation}
	\dd s^2 = g_{ij} \lambda^i \lambda^j \, ,\qquad e^a = e^a_i \lambda^i\,.
\end{equation}
We call $E_a^i$ the inverse vielbein, $E_a^i e^b_i= \delta_a^b$. 

We use $g_{ij}$ to lower the $i$ indices; $\delta_{ab}$ lowers the $a$ indices, as usual. Imposing the invariance of $g_{ij}$ (so that it makes sense on $G/H$) gives, recalling again \cref{eq:Lrl},
\begin{equation}
	g_{kj}f^k{}_{i \alpha}+g_{ik}f^k{}_{j \alpha}= 2f_{(ij)\alpha}=0\,.
\end{equation}
We also freely use the vielbein to convert indices from $i$ to $a$ and vice versa; for example, $f^a{}_{bc}\equiv e^a_i E_b^j E_c^k f^i{}_{jk}$. With this notation, we can write
\begin{equation}
	\dd e^a = -\frac12 f^a{}_{bc} e^b \wedge e^c -f^a{}_{\alpha c} \lambda^\alpha \wedge e^c\,.
\end{equation}
This is the action of $\dd$ on $G$; the action of $\dd_M$, the exterior differential on $M=G/H$, can be obtained by replacing $\lambda^\alpha\to (h^{-1}Ah)^\alpha$, in view of (\ref{eq:l-A}). Recalling the first structure equation, $\dd_M e^a + \omega^{ab} \wedge e_b=0$, this leads to an expression for the spin connection on $M$:
\begin{equation}\label{eq:spin-conn}
	\omega^{ab}= - G^{ab}{}_c e^c - f^{ab}{}_\alpha (h^{-1} A h)^\alpha
	\, ,\qquad G_{abc}\equiv \frac12 (f_{abc}-f_{bac}+f_{cba})\,.
\end{equation}
Using now the second structure equation, $\dd_M \omega^{ab}+ \omega^a{}_c \wedge \omega^{cb}=0$, a lengthy calculation leads to
\begin{equation}\label{eq:riemann}
	R_{abcd}= G_{abe}f^e{}_{cd} + 2G_a{}^e{}_{[c|} G_{eb|d]}+ f_{ab\alpha} f^\alpha{}_{cd}\,.
\end{equation}
The Ricci tensor is then $R_{ab}=e_a^i e_b^j R_{ij}$, with
\begin{equation}\label{eq:Rij}
	R_{ij}= -\frac12 f^k{}_{il} f_{kj}{}^l -\frac12 f^k{}_{il}f^l{}_{jk}
	- f^k{}_{(i|\alpha} f^\alpha{}_{j)k} +\frac14 f_{ikl} f_j{}^{kl}+f^{kl}{}_k f_{(ij)l}\,. 
\end{equation}

\subsection{Gauge fields from coset reductions}
\label{ssec:coset-reduction}

We now look for solutions to the heterotic $\mathrm{SO}(16)\times \mathrm{SO}(16)$ string on
\begin{equation}
	{\rm AdS}_{10-d}\times G/H\,.
\end{equation}
To illustrate the general strategy, we take the internal metric as 
\begin{equation}\label{eq:R2dij}
	g_{ij}= R^2 \delta_{ij}\,.
\end{equation}
We have looked for more general solutions in the individual AdS$_4$ examples below, but we have found that they all reduce, in fact, to \cref{eq:R2dij}.

The three-form and the gauge field can be taken proportional to those in \cref{eq:3form0,eq:hFh},
\begin{equation}\label[pluralequation]{eq:HF-ans}
	H= - \frac16 h R^3 \tilde f_{ijk} \lambda^{ijk} \, ,\qquad
	(h^{-1}F h)^\alpha = -\frac12 f f^\alpha{}_{jk} \lambda^{jk}\,.  
\end{equation}
The three-form Bianchi identity from \cref{eq:het-eom-gauge} is then satisfied with
\begin{equation}\label{eq:hf}
	h R^3 = \frac{a}{2} f^2 \,,\qquad a \equiv \frac{k_{16} \ell_s^2}{60}\,,
\end{equation}
where $k_{16}$ was introduced after \cref{eq:trTr}.

Before proceeding, we want to comment on \cref{eq:R2dij}. Taking the metric proportional to the identity, we consider only a subset of the possible vacua. One can try to approach the problem in its full generality, but the Einstein equations (\cref{ssec:coset-reduction-einstein} below) become less tractable. In fact, leaving the metric implicit results in substantially more complex equations. These can be simplified because $H$ acts on the coset, and therefore one can split $G/H$ into $H$-modules $V_r$. Assuming that each $V_r$ appears with multiplicity $1$, one can choose a basis so that the most general left-invariant metric is proportional to the identity on each $V_r$, $g_{ij}|_{V_r}=R_r^2 \delta_{ij}|_{V_r}$.
This leads to a system of equations for the radii $R_r$ that, in principle, fixes their values. However, we have not found a way to solve these equations for $R_r$ in full generality: they contain different combinations of the structure constants with indices restricted to different $H$-modules $V_r$, and the analysis soon becomes dependent on the explicit coset that one is considering.
Since the metric of \cref{eq:R2dij} is sufficient for all the examples of \cref{ssec:ads-examples}, we focus on that case only, leaving the most general setup for future investigations.

\subsubsection{Einstein equations}\label{ssec:coset-reduction-einstein}

Recall that in our conventions we lower the $i$ indices with the metric and the $\alpha$ indices with $\delta_{\alpha \beta}$. The $\tilde f_{ijk}= \frac1{R^2}f_{ijk}$ are totally antisymmetric; on the other hand, $f_{ij \alpha}= R^2 \tilde f_{ij \alpha}= R^2 \tilde f_{\alpha ij}= R^2 f_{\alpha ij}$. The Ricci tensor in \cref{eq:Rij} simplifies to
\begin{equation}\label{eq:Rij-R2d}
	R_{ij}= \frac14 f_{ikl}f_j{}^{kl} - f^k{}_{i\alpha} f^\alpha{}_{jk}\,.
\end{equation}
The form contributions are easily computed: 
\begin{eq}\label[pluralequation]{eq:iHiH-iFiF}
    \iota_i H \cdot \iota_j H = \frac{(hR)^2}{2} f_{ikl}f_j{}^{kl}\mcomma \qquad \iota_i F_\alpha \cdot \iota_j F^\alpha= \frac{f^2}{R^2} f_{\alpha ik} f^\alpha{}_j{}^k \mperiod
\end{eq}
The two structures appearing in \cref{eq:Rij-R2d} can be related using the Killing form in \cref{eq:killing}:
\begin{equation}\label{eq:kAd-ia}
	-k \delta_{ij} = f^k{}_{il} f^l{}_{jk} + 2 f^k{}_{i \alpha} f^\alpha{}_{jk} \quad \Rightarrow \quad 
	\frac{k d}{R^2}= f_{ijk}f^{ijk}+ 2 f_{ij \alpha}f^{ij\alpha}\,.
\end{equation}

Using \cref{eq:Rij-R2d,eq:iHiH-iFiF}, the Einstein equations reduce to 
\begin{eq}\label{eq:Einstein-coset}
    \gamma f_{ikl} f_j{}^{kl} + \left(\alpha + \beta R^2 f_{klm}f^{klm}\right) \delta_{ij}=0\mcomma
\end{eq}
with 
\begin{eq}
    \alpha=\frac{k}2 \left[1+ \left(\frac{d}{20}-1\right)hR\right]\mcomma \qquad \beta=\frac{h^2R^2}{60}- \frac{hR}{40}\mcomma \qquad \gamma=-\frac14 \left(h R -1 \right)^2\mperiod
\end{eq}
Combining \cref{eq:Einstein-coset} with its trace gives
\begin{equation}\label{eq:gq}
	\gamma \left(f_{ikl} f_j{}^{kl} + \frac{\alpha}{\gamma + d \beta}\delta_{ij}\right)=0\,.
\end{equation}
Setting to zero the parenthesis requires that $f_{ikl} f_j{}^{kl}$ be proportional to the identity, 
\begin{equation}\label{eq:kM}
	f_{ikl} f_j{}^{kl}= k_M \delta_{ij}\,.
\end{equation}
This is a non-trivial requirement on $G/H$, but it is often realized. We saw in \cref{eq:q-inv} that the left-hand side of \cref{eq:kM} is an invariant symmetric tensor. For several coset spaces, there is only one such tensor up to rescaling, so \cref{eq:kM} is guaranteed to hold. We will see that other non-trivial examples exist, where there are several independent invariant symmetric tensors, and yet \cref{eq:kM} is satisfied. 

Combining \cref{eq:kM} with \cref{eq:gq} gives, assuming $k_M\neq 0$,
\begin{eq}\label{eq:AdS-coset-radius}
    hR = \frac{3(d-20)(k-k_M)\pm \sqrt{9 k^2 (d-20)^2+3k_M d(40-3d)(2k-k_M)}}{4(15-d)k_M}\mperiod
\end{eq}
For a fixed dimension $d$, this only depends on the ratio $k/k_M$. Moreover, \cref{eq:kAd-ia} implies that $k\ge k_M$, and the trace of \cref{eq:kM} shows that $k_M\ge 0$. Therefore, one and only one solution from \cref{eq:AdS-coset-radius} is positive and real (which is important in view of \cref{eq:hf}).

In principle, \cref{eq:gq} allows the second branch $\gamma=0$, which implies $h=1/R$. Among the spaces $G/H$ with dimension $d=6$, we did not find any example where $\gamma=0$; in fact, this is inconsistent with \cref{eq:kM} and the inequality $k\geq k_M$. Note that for this reason, the two branches cannot possibly intersect. 

The particular case $k_M=0$ deserves a separate treatment. The trace of \cref{eq:kM} shows that $f_{ijk}=0$, so that $H=0$. \Cref{eq:3form0} is still valid, and now $F_\alpha \wedge F^\alpha=0$. The rescaling parameter $f$ is no longer fixed by the three-form Bianchi identity as in \cref{eq:hf}. Repeating the analysis above for \cref{eq:gq} gives
\begin{equation} \label{eq:kM=0-case}
    \frac{ a f^2}{R^2} = \frac{40}{20-d}\,.
\end{equation}
Formally, these are valid solutions to our original system of \cref{eq:het-eom-gauge}. However, the fact that ${\rm tr}F\wedge F=0$ means that we can no longer ignore the Riemann$^2$ contributions to the three-form Bianchi identity. Therefore, we cannot trust the solutions in \cref{eq:kM=0-case}. Another perspective is that flux quantization in such solutions typically fails to fix all continuous parameters, thus leaving some free moduli; the latter are not protected by supersymmetry, so they are expected to be lifted by string corrections.
Thus, we refrain from discussing \cref{eq:kM=0-case} further.

\subsubsection{Form equations of motion}

We first consider $\dd * H =0$. We provide two arguments to show that this is automatically satisfied for the three-form of \cref{eq:HF-ans}. The first relies on rewriting $* \, \dd *H= \dd^\dagger H= - \frac12 \nabla^k H_{kij} \lambda^{ij}$. In terms of the vielbein and the spin connection, in general $\nabla_a H_{bcd}= \partial_a H_{bcd}+ 3 \omega_{[b|i}{}^e E_a^i H_{|cd]e}$. For coset spaces, we can take $\omega$ from \cref{eq:spin-conn}. The terms containing $A$ in $\nabla_a H_{bcd}$ are proportional to $f_{[i|}{}^l{}_\alpha H_{l|jk]}$, which vanishes because $H$ from \cref{eq:HF-ans} is an invariant form. The remaining terms in $\nabla^k H_{kij}=0$ give
\begin{equation}\label{eq:d*H-1}
	f_{kl[i} f_{j]}{}^{kl}=0\,,
\end{equation}
which is automatically satisfied.

Alternatively, we can use the definition of the Hodge star on $G/H$, which in our case gives $* \lambda^{i_1\ldots i_k}= \frac{R^{d-2k}}{(d-k)!} \epsilon_{i_{k+1}\ldots i_d}{}^{i_1\ldots i_k} \lambda^{i_{k+1}\ldots i_d}$. When evaluating $\dd *H$, \cref{eq:dli} generates two terms; the term containing $\lambda^\alpha$ is proportional to 
\begin{equation}
	\tilde f_{ijk} \epsilon_{i_4[i_5 \ldots i_d}{}^{ijk}f^{i_4}{}_{l]\alpha}\,.
\end{equation}
Contracting this with a further $\epsilon^{i_5\ldots i_d l}{}_{j_1j_2j_3}$, we obtain $f^\alpha{}_{l[j_1}f^{l}{}_{j_2j_3]}$, which vanishes by \cref{eq:jacobi-red}.\footnote{We have obtained that $*H$ is an invariant form, \cref{eq:a-inv}: $\dd *H$ does not contain a $\lambda^\alpha$ term. This was expected, given that both $H$ and the metric are invariant.} The remaining term in $\dd *H$ gives
\begin{equation}\label{eq:d*H-2}
	* \, \dd * H = \frac{(-1)^d}{2} h R^3 \tilde f_{ijk} f_l{}^{ij} \lambda^{lk}=0\,,
\end{equation}
which agrees with our previous computation in \cref{eq:d*H-1}.

We now consider the equation of motion for the gauge field. We rewrite it using a trick similar to \cref{eq:d(hFh)}:
\begin{equation}\label{eq:Feq}
    h^{-1} (\dd * F + [A, * F]+ F \wedge *H) h=  \dd (h^{-1} *F h ) + [\lambda^\alpha T_\alpha,h^{-1} *F h] + h^{-1} F h \wedge *H\,.
\end{equation}
The computation now follows the same steps as the one above, which led to \cref{eq:d*H-2}.
Recalling \cref{eq:HF-ans}, the first term on the right-hand side of \cref{eq:Feq}, $\dd (h^{-1} *F h )$, generates two sub-terms from \cref{eq:dli}; one of these cancels with $[\lambda^\beta T_\beta,h^{-1} *F h]= T_\alpha f^\alpha{}_{\beta \gamma} \lambda^\beta \wedge h^{-1} *F^\gamma h $ using the identity
\begin{equation}
	(d-2) \epsilon_{k i_4 \ldots i_d}{}^{ij} f^\alpha{}_{ij} f^k{}_{\gamma l}+ \epsilon_{li_4\ldots i_d}{}^{ij} f^\alpha{}_{\beta \gamma} f^\beta{}_{ij}=0\,,
\end{equation}
which can be obtained by contracting with $\epsilon^{l i_4\ldots i_6}{_{j_1j_2}}$ and using one of the Jacobi identities in \cref{eq:jacobi-red-a}. The remaining terms assemble into 
\begin{equation}
		*\left(\dd (h^{-1} *F h ) + [\lambda^\alpha T_\alpha,h^{-1} *F h] + h^{-1} F h \wedge *H\right)=
		\frac{f}{2} R^2(1+ h R)f^{\alpha jk} \tilde f_{ijk} \lambda^i T_\alpha\,.
\end{equation}
The Killing form is diagonal by assumption; therefore, $0=f^J{}_{\alpha K}f^K{}_{iJ}=f^j{}_{\alpha k} f^k{}_{ij}$. It then follows that \cref{eq:Feq} vanishes.

\subsubsection{Corrections}

We promised at the beginning of this section that we would work in a regime where the gauge curvature is large, so that the often quoted curvature corrections to the heterotic equations of motion and Bianchi identities would be negligible. 
We now argue that this is indeed the case for the solutions that we have just found.
Recall that this is unlike the usual regime for Minkowski compactifications of the supersymmetric heterotic string, where the gauge field terms are small and comparable to the first curvature corrections.

The $\lambda^i$ are dimensionless; they can be expressed in terms of coordinates $\theta^i$ with a fixed range. The physical coordinates $x^m$, which have dimensions of length, have a different index $m$. The metric is $\dd s^2= g_{mn} \dd x^m \dd x^n= g_{ij} \lambda^i \lambda^j$; therefore, $g_{mn}$ is dimensionless while $g_{ij}$ has dimensions of length$^2$, as we saw in \cref{eq:R2dij}. 

With this clarification, in terms of the overall radius $R$, we see from \cref{eq:Rij} that  $R_{ij}\sim R^0$. From \cref{eq:HF-ans}, both $\iota_i H \cdot \iota_j H$ and $H^2 g_{ij}$ carry dimensions of $(h R^3)^2 R^{-4}= h^2 R^2$. Finally, both ${\rm tr}(\iota_i F \cdot \iota_j F)$ and ${\rm tr}F^2 g_{ij}$ have dimensions of $\ell_s^2 f^2 R^{-2}$. Using \cref{eq:hf,eq:AdS-coset-radius}, we see that all the terms in \cref{eq:intR}---after converting the indices to $i$, $j$---are of the same order (as they should), namely $R^0$.

The leading correction to \cref{eq:intR} is proportional to 
\begin{equation}\label{eq:R+2}
    R^+_{mpqr}R_n^{+pqr}\,,
\end{equation}
where  $R^+_{mnpq}= R_{mnpq} - \nabla_{[p} H_{q]mn}-\frac12 H_{mr[p} H_{q]n}{}^r$ is the curvature of $(\Gamma+\frac12 H)^m{}_{np}$.
Converting the $a$ indices to $i$ in \cref{eq:riemann} gives $R_{ijkl}\sim R^2$. So $R_{ii_1i_2i_3} R_j{}^{i_1 i_2 i_3}\sim R^{-2}$. The terms of the form $R\nabla H$ and $HH$ in \cref{eq:R+2} are of the same order. Therefore, these are parametrically smaller than the terms already present in \cref{eq:intR}. 

The parameter $R$ is not a modulus: it is quantized by demanding that the gauge field Chern classes $c_k$ be integers. Control is achieved when the $c_k$ are large: for example, $\int {\rm tr}(F\wedge F)\sim f^2 \sim \ell_s^{-2}R^2$. As we anticipated, this regime is quite different from the standard embedding used in Minkowski compactifications of the supersymmetric heterotic string, where famously the Chern classes of the gauge bundle are fixed to be equal to those of $TM$; in other words, in that case the $F\wedge F$ and $R\wedge R$ terms are of the same order and must both be taken into account, whereas in our situation $F\wedge F$ dominates over the $R\wedge R$ terms.

\subsection{\texorpdfstring{Examples of AdS$_4$ vacua}{Examples of AdS4 vacua}}
\label{ssec:ads-examples}

We now provide more details for some particular $G/H$ of dimension six, with external space AdS$_4$. This is both to make our strategy more concrete, to work out flux quantization, and to check that there are no solutions beyond the Ansatz of \cref{eq:R2dij}.

Compact, simply connected coset spaces $G/H$ with $G$ simple were classified in \cite{klaus-homogeneous}.\footnote{Another natural option might have been solvmanifolds, but by \cite[Cor.~4.4]{gromov-lawson1} and \cite[Thm.~A]{gromov-lawson3} they do not admit metrics of positive scalar curvature, which is required by \cref{eq:intR}.} In six dimensions, denoting homeomorphisms by $\cong$ (see also \cite{Chapline:1982wy}), we have
 \begin{itemize}
    \item ${\rm SU}(3)/{\rm U}(1)^2\cong$ the flag manifold $\mathbb{F}(1,2;3)$;
    \item ${\rm Sp}(2)/({\rm Sp}(1)\times {\rm U}(1))\cong \mathbb{CP}^3$;
    \item $G_2/{\rm SU}(3)\cong S^6$;
    \item ${\rm SO}(7)/{\rm SO}(6)\cong S^6$;
    \item ${\rm SU}(4)/{\rm U}(3)\cong \mathbb{CP}^3$;
    \item 
    ${\rm SO}(5)/({\rm SO}(3)\times{\rm SO}(2))\cong$ the real Grassmannian $G_{2,5}$ ($\cong$ a complex quadric in $\mathbb{CP}^4$).
\end{itemize}
Note that none of these has $H^3\neq 0$, in agreement with our comments at the end of \cref{ssec:Tpq}.

\subsubsection{\texorpdfstring{AdS$_4\times\mathbb{F}(1,2;3)$}{AdS4 x flag manifold}}

The flag manifold $\mathbb{F}(1,2;3)$ is the space of nested lines and planes in $\C^3$. It is the coset ${\rm SU}(3)/{\rm U}(1)^2$, so that the Kaluza--Klein approach involves two ${\rm U}(1)$ gauge fields embedded in ${\rm SO}(16)\times {\rm SO}(16)$.

Rather than applying the procedure of \cref{ssec:coset-reduction} step by step, we construct the solution directly from the left-invariant fields on the coset, which may, in principle, yield a more general result. We will find that even considering the most general left-invariant metric on the coset and the most general form fields compatible with their equations of motion and Bianchi identities, the equations will uniquely select the solution of \cref{ssec:coset-reduction}. We will comment on this result in \cref{ssec:comments}. 

We choose the structure constants of ${\rm SU}(3)$ as
\begin{eq}
    &{f^1}_{54}={f^1}_{36}={f^2}_{46}={f^2}_{35}={f^3}_{47}={f^5}_{76}=1\mcomma \qquad {f^1}_{27}=2\mcomma \\
    & {f^3}_{48}={f^5}_{68}=\sqrt{3} \mcomma \qquad \text{and cyclic.}
\end{eq}
These are obtained by taking the usual Gell-Mann matrices, divided by a factor of $i$ and reordered so that the ${\rm U}(1)^2$ subgroup corresponds to the $7$ and $8$ directions.
A basis of left-invariant two-forms and three-forms is given by
\begin{eq}\label[pluralequation]{eq:flag-2-forms}
    j_1= \lambda^{12}\mcomma\qquad j_2=- \lambda^{34}\mcomma\qquad j_3= \lambda^{56}\mcomma
\end{eq}
and 
\begin{eq}\label[pluralequation]{eq:flag-3-forms}
    \psi=\lambda^{135}+\lambda^{146}-\lambda^{236}+\lambda^{245}\mcomma \qquad \Tilde{\psi}=-\lambda^{136}+\lambda^{145}-\lambda^{235}-\lambda^{246}\mperiod
\end{eq}
These satisfy 
\begin{eq}
    &-\psi+i\Tilde{\psi}=(\lambda^1+ i \lambda^2)\wedge (-\lambda^3 + i \lambda^4)\wedge (\lambda^5 +i \lambda^6)\mcomma \\
    & \frac{1}{6}(j_1+j_2+j_3)^3=\frac{i}{8}(-\psi+i\Tilde{\psi})\wedge\overbar{(-\psi+i\Tilde{\psi})}\mcomma \qquad \dd j_i=\psi \mcomma \\
    &  j_i\wedge \psi=j_i\wedge \Tilde{\psi}=0\mcomma \qquad \dd\Tilde{\psi}=4(j_1\wedge j_2+j_1\wedge j_3+j_2\wedge j_3) \mperiod
\end{eq}
We will use these as building blocks for $H$ and $F$.
The most general left-invariant metric on the flag manifold is, in the coframe basis, 
\begin{eq}
    \dd s^2=\alpha_1^2 (\lambda^1\lambda^1+\lambda^2\lambda^2)+\alpha_2^2 (\lambda^3\lambda^3+\lambda^4\lambda^4)+\alpha_3^2 (\lambda^5\lambda^5+\lambda^6\lambda^6)\mperiod
\end{eq}
We use it to define the following two combinations $J$ and $\Omega$: 
\begin{eq}
    J&=J_1+J_2+J_3=\alpha_1^2 j_1+\alpha_2^2 j_2+\alpha_3^2 j_3\mcomma\\
    \Omega &=\alpha_1\alpha_2\alpha_3(-\psi+i\Tilde{\psi})\mperiod
\end{eq}

The most general choice of gauge fields $F^{\alpha=1,2}$ in the ${\rm U}(1)^2$ subgroup that solves the Bianchi identities is  
\begin{eq}\label{eq:flag-F2}
    F^\alpha=A^\alpha (j_1-j_2)+B^\alpha (j_2-j_3)\mcomma
\end{eq}
where $A^\alpha$ and $B^\alpha$ are constants written in vector notation, and $j_i$ are the left-invariant two-forms of \cref{eq:flag-2-forms}.
Similarly, imposing the equations of motion of the three-form leads to an expression for $H$ in terms of the left-invariant form $\Tilde{\psi}$:
\begin{eq}\label{eq:flag-H3}
    H=h \text{Im}\Omega=h\alpha_1\alpha_2\alpha_3\Tilde{\psi}\mcomma 
\end{eq}
where $h$ is a constant.
The equations of motion of the forms are equivalent to
\begin{eq}\label[pluralequation]{eq:flag_Bianchi}
    \frac4a \alpha_1\alpha_2\alpha_3 h =A^2-A\cdot B=B^2-A\cdot B=A\cdot B\mcomma
\end{eq}
where $a$ is defined in \cref{eq:hf}.
\Cref{eq:flag_Bianchi} are solved by 
\begin{eq}\label[pluralequation]{eq:flag-f-eq}
    A=f\begin{pmatrix}-2\\0\end{pmatrix}\mcomma \qquad B=f\begin{pmatrix}-1\\\sqrt{3}\end{pmatrix}\mcomma \qquad \text{where } a f^2 =2\alpha_1\alpha_2\alpha_3h\mperiod
\end{eq}
In fact, \cref{eq:flag-F2,eq:flag-H3} match the gauge fields in \cref{eq:HF-ans}, when written in terms of the structure constants, together with the condition in \cref{eq:hf}.
Note that $F$ can be consistently quantized because the lengths of the two $\rm{U}(1)$ fibers are unequal. For both $\alpha=1,2$ and both two-cycles $C_I$, we need to impose $\frac1{\Delta \psi_\alpha}\int_{C_I} F^\alpha \in \mathbb{Z}$; the generators are $T_7 = \frac1{i} {\rm diag} (1,-1,0)$ and $T_8= \frac1{\sqrt 3 i} {\rm diag}(1,1,-2)$, from which we see that $\Delta \psi_1= 2\pi$ and $\Delta \psi_2= 2\sqrt3 \pi$. 
We keep the parameter $f$ instead of the flux number $n\in\Z$ to avoid cumbersome expressions, but we replace it with $n$, using $f\propto n$, in the final expressions. There is no quantization condition on $H$ because there are no closed three-cycles.

The metric equations are
\begin{eq}
    &\frac{3}{5}h^2+\frac{2}{5}h\left(9\frac{\alpha_2\alpha_3}{\alpha_1^3}-\frac{\alpha_3\alpha_1}{\alpha_2^3}-\frac{\alpha_1\alpha_2}{\alpha_3^3}\right)+\frac{-\alpha_1^4+\alpha_2^4+\alpha_3^4-6\alpha_2^2\alpha_3^2}{\alpha_1^2 \alpha_2^2 \alpha_3^2}\mcomma \\
    & \text{and cyclic.}
\end{eq}
After taking linear combinations, one can show that this system admits 8 solutions, of which 4 are compatible with the positive sign of $\alpha_1\alpha_2\alpha_3 h$ required by \cref{eq:flag-f-eq}. The solutions can be expressed as
\begin{eq}
    (\alpha_1,\alpha_2,\alpha_3)=(R,R,R),(R,-R,-R),(-R,-R,R),(-R,R,-R)\mcomma
\end{eq}
where $R$ satisfies
\begin{eq}
    hR=\frac{2\sqrt{31}-7}{3}\mperiod
\end{eq}
Therefore, up to signs that do not affect any physical quantities, there is only one solution with all the parameters $\alpha_i$ equal to $R>0$. We have collapsed on the type of solutions of \cref{ssec:coset-reduction}.
An interesting consequence is that $J$ and $\Omega$ now define a nearly-K\"{a}hler structure on the flag manifold:
\begin{eq}
    J\wedge\Omega=0\mcomma \qquad \frac{1}{6}J^3=\frac{i}{8}\Omega\wedge\Bar{\Omega}\mcomma \qquad \dd J=-3 \text{Re}\Omega\mcomma \qquad \dd\text{Im}\Omega=2 J^2\mperiod
\end{eq}
We will comment on this in \cref{ssec:comments}.

The equations with $\alpha_i=R$ are the same as in \cref{ssec:coset-reduction-einstein}. The metric and the dilaton equations become
\begin{eq}\label[pluralequation]{eq:flag-metric-dilaton}
    &\frac{5}{R^2}-\frac{3}{5}h^2-\frac{14}{5}\frac{h}{R} =0 \qquad && \Rightarrow \ hR  =\frac{2\sqrt{31}-7}{3}\mcomma \\
    &\frac{1}{5}h^2 +\frac{h}{10R} =T e^{2\phi_0} \qquad && \Rightarrow \ e^{2\phi_0}\propto (T R^2)^{-1}\mcomma
\end{eq}
thus reproducing \cref{eq:AdS-coset-radius} with $d=6$ and $k/k_M=3$; this is, in fact, the correct value for ${\rm SU}(3)/{\rm U}(1)^2$.

The complete solution depends on the single free parameter $n$, the quantized flux number, and the relevant quantities scale as
\begin{eq}\label{eq:flag-scaling}
    L\sim R\sim T^{-\frac{1}{2}} e^{-\phi_0}\sim n\mcomma
\end{eq}
ensuring that the solution is reliable for large values of the quantized flux.

\subsubsection{\texorpdfstring{AdS$_4\times\C\mathbb{P}^3$}{AdS4 x CP3}}
\label{ssub:cp3}

The second example uses the complex projective space $\C\mathbb{P}^3$ as the internal component, realized as the coset ${\rm Sp}(2)/({\rm Sp}(1)\times {\rm U}(1))$ in which $\mathfrak{su}(2)\oplus\mathfrak{u}(1)$ is embedded into an $\mathfrak{su}(2)\oplus\mathfrak{su}(2)\cong\mathfrak{so}(4)$ subalgebra of $\mathfrak{sp}(2)\cong\mathfrak{so}(5)$.
The approach of \cref{ssec:coset-reduction} thus involves gauge fluxes in an $\rm{SU}(2)\times\rm{U}(1)$ subgroup of $\mathrm{SO}(16)\times \mathrm{SO}(16)$.

We choose the structure constants so that $\rm{SU}(2)\times\rm{U}(1)$ is included in the last 4 indices,
\begin{eq}
    &{f^5}_{41}={f^5}_{32}={f^6}_{13}={f^6}_{42}=\frac{1}{2}\mcomma \qquad {f^7}_{56}={f^{10}}_{89}=-1\mcomma \\
    & {f^7}_{21}={f^7}_{43}={f^8}_{14}={f^8}_{32}={f^9}_{13}={f^9}_{24}={f^{10}}_{34}={f^{10}}_{21}=\frac{1}{2} \mcomma \qquad \text{and cyclic.}
\end{eq}
We define a basis of left-invariant two-forms,
\begin{eq}
    j_B= \frac{1}{8}(\lambda^{12}+\lambda^{34})\mcomma \qquad j_F = -\lambda^{56} \mcomma
\end{eq}
and a basis of left-invariant three-forms,
\begin{eq}
    \psi=\frac{1}{8}(-\lambda^{135}-\lambda^{146}-\lambda^{236}+ \lambda^{245})\mcomma \qquad \Tilde{\psi}=\frac{1}{8}(-\lambda^{136}+\lambda^{145}+\lambda^{235}+\lambda^{246})\mcomma
\end{eq}
which satisfy
\begin{eq}
    & \dd j_B=\frac{1}{4}\dd j_F=\psi\mcomma \qquad \dd\psi=0\mcomma \qquad \dd\Tilde{\psi}=8j_B\wedge j_B+2j_F\wedge j_B\mcomma \\
    & \psi\wedge\Tilde{\psi}=-2j_B\wedge j_B\wedge j_F\mcomma \qquad \dd(j_{B,F}\wedge j_{B,F})=0\mperiod
\end{eq}
The most general left-invariant metric in the coframe basis is
\begin{eq}
    \dd s^2 =\frac{R^2}{4}\left[\frac{1}{\sigma}(\lambda^1 \lambda^1+\lambda^2 \lambda^2+\lambda^3 \lambda^3+\lambda^4 \lambda^4)+\lambda^5 \lambda^5+\lambda^6 \lambda^6\right]\mperiod
\end{eq}
We use it to define the following two combinations $J$ and $\Omega$:
\begin{eq}
    J\equiv J_B+J_F\equiv R^2\left(\frac{2}{\sigma}j_B+\frac{1}{4}j_F\right)\mcomma \qquad \Omega=\frac{R^3}{\sigma}(-\psi + i \Tilde{\psi})\mperiod
\end{eq}

Again, the most general gauge fields that are compatible with the Bianchi identities are those that are generated by the strategy in \cref{ssec:coset-reduction}. In our conventions, the generators of ${\text{SU(2)}\times\text{U(1)}}$ are identified with $\lambda^{8,9,10}$ and $\lambda^7$, so that the gauge fields of \cref{eq:HF-ans} are
\begin{eq}
    F_{\text{U(1)}}&=f_1(4j_B-j_F)\mcomma \\
    F_{\text{SU(2)}}&=\frac{f_2}{2}\left(-\lambda^{14}+\lambda^{23},-\lambda^{13}-\lambda^{24},\lambda^{12}-\lambda^{34}\right)\mperiod
\end{eq}
Similarly, the Kalb--Ramond field strength is
\begin{eq}
    H=h\frac{\text{Im}\Omega}{2}=h\frac{R^3}{2\sigma}\Tilde{\psi}\mperiod
\end{eq}
The equations of motion of the gauge fields and the Bianchi identity of the three-form flux are all satisfied provided that
\begin{eq}
    f_1=f_2\equiv f \qquad \text{and}\qquad a f^2=\frac{R^3}{4\sigma} h\mcomma
\end{eq}
where $a$ is defined in \cref{eq:hf}.

The Einstein equations are then equivalent to
\begin{eq}
    &\frac{(6-\sigma)\sigma}{R^2}=\frac{3}{20}h^2+\frac{4h}{R\sigma}\left(\frac{2}{5}\sigma^2-\frac{1}{20}\right)\mcomma \\
    &\frac{2(\sigma-1)}{R^2\sigma}\left[\sigma(\sigma-2)+hR(\sigma+1)\right]=0\mperiod
\end{eq}
The only solution that is compatible with the positive sign of $\sigma$ is $\sigma=1$, in which case $J$ and $\Omega$ define a nearly-K\"{a}hler structure and 
\begin{eq}
    h\frac{R}{2}=\frac{2\sqrt{31}-7}{3}\mperiod
\end{eq}
This matches \cref{eq:AdS-coset-radius} with $d=6$ and $k/k_M=3$, which is the appropriate value for ${\rm Sp}(2)/({\rm Sp}(1)\times {\rm U}(1))$.
The remaining dilaton equation fixes $\phi_0$, leading to the same scaling as in the second of \cref{eq:flag-metric-dilaton}.
The physical parameters scale with the flux number $n$ in the same way as in \cref{eq:flag-scaling}.

\subsubsection{\texorpdfstring{AdS$_4\times S^6$}{AdS4 x S6}}

The last solution that we present in detail is based on the coset $G_2/{\rm SU}(3)$, which is topologically a six-sphere $S^6$.
We fix the structure constants of $G_2$,
\begin{eq}
    {f^1}_{63}&={f^1}_{45}={f^2}_{53}={f^2}_{64}=\frac{1}{\sqrt{3}}\mcomma \qquad {f^{14}}_{43}={f^{14}}_{56}=\frac{1}{2\sqrt{3}}\mcomma\qquad {f^{14}}_{21}=\frac{1}{\sqrt{3}} \mcomma  \\
    {f^7}_{36}&={f^7}_{45}={f^8}_{53}={f^8}_{46}={f^9}_{56}={f^9}_{34}={f^{10}}_{16}={f^{10}}_{52}\\
    &={f^{11}}_{51}={f^{11}}_{62}={f^{12}}_{41}={f^{12}}_{32}={f^{13}}_{31}={f^{13}}_{24}=\frac{1}{2}\mcomma
\end{eq}
together with a copy of SU(3) in the last 8 indices, with 
\begin{eq}
    {f^{i+6}}_{j+6,k+6}={{f^{(\text{SU(3)})}}^i}_{jk}\mcomma
\end{eq}
where 
\begin{eq}
    &{{f^{(\text{SU(3)})}}^1}_{65}={{f^{(\text{SU(3)})}}^1}_{47}={{f^{(\text{SU(3)})}}^2}_{57}={{f^{(\text{SU(3)})}}^2}_{46}={{f^{(\text{SU(3)})}}^4}_{53}={{f^{(\text{SU(3)})}}^6}_{37}=\frac{1}{2}\mcomma \\
    &{{f^{(\text{SU(3)})}}^1}_{23}=1\mcomma\qquad {{f^{(\text{SU(3)})}}^4}_{58}={{f^{(\text{SU(3)})}}^6}_{78}=\frac{\sqrt{3}}{2}\mperiod
\end{eq}
We define a basis of left-invariant two- and three-forms,
\begin{eq}
    j&=\lambda^{12}-\lambda^{34}+\lambda^{56}\mcomma \\
    \psi&=-\lambda^{135}-\lambda^{146}+\lambda^{236}-\lambda^{245}\mcomma \\
    \Tilde{\psi}&=\lambda^{136}-\lambda^{145}+\lambda^{235}+\lambda^{246}\mcomma
\end{eq}
which satisfy
\begin{eq}
    \dd j=\sqrt{3}\psi\mcomma \qquad \dd\psi=0\mcomma \qquad \dd\Tilde{\psi}=\frac{2}{\sqrt{3}} j^2\mcomma\qquad \psi\wedge\Tilde{\psi}=-\frac{2}{3}j^3\mperiod
\end{eq}
The most general left-invariant metric in the coframe basis is proportional to the identity,
\begin{eq}
    \dd s^2 = R^2 \sum_i \lambda^i\lambda^i\mcomma
\end{eq}
so that the analysis of \cref{ssec:coset-reduction} applies without any changes.
As in the previous cases, the forms
\begin{eq}
    J=\frac{R^2}{3} j\mcomma\qquad \Omega=\frac{R^3}{3\sqrt{3}} (-\psi+i\Tilde{\psi})
\end{eq}
define a nearly-K\"{a}hler structure on the sphere.

\Cref{ssec:coset-reduction} instructs us to turn on SU(3) gauge fields such that the SU(3) components of the gauge connection are 
\begin{eq}
    A^a= f \lambda^{a+6}\mperiod
\end{eq}
For the Kalb--Ramond field strength, we must take
\begin{eq}
    H=h R^3 \frac{\Tilde{\psi}}{\sqrt{3}}\mperiod
\end{eq}
The equations of motion and the Bianchi identities for $H$ and $F$ reduce to \cref{eq:hf}, while the Einstein equations become
\begin{eq}
    \frac{1}{5}(hR)^2+\frac{14}{15}hR-\frac{5}{3}=0 \ \Rightarrow \ hR=\frac{2\sqrt{31}-7}{3}\mperiod
\end{eq}
This reproduces \cref{eq:AdS-coset-radius} with $d=6$ and $k/k_M=3$, which is again the appropriate value for $G_2/{\rm SU}(3)$.
The dilaton equation fixes $\phi_0$ as in the other cases. 
The gauge flux is quantized despite the absence of two-cycles, because the integral of the third Chern character $\frac{1}{3! (2\pi)^3}{\rm tr}\left(F\wedge F \wedge F\right)$ on the six-manifold is non-trivial (${\rm SU}(3)$ has a non-vanishing totally symmetric invariant three-tensor, $d_{abc}\neq0$, which enters the cubic Casimir). The scaling of all relevant quantities in terms of the flux number $n$ is
\begin{eq}
    L\sim R\sim T^{-\frac{1}{2}} e^{-\phi_0}\sim n^{\frac{1}{3}}\mperiod
\end{eq}

\subsubsection{The remaining cases}

An analogous solution exists on $S^3\times S^3$, seen as the group ${\rm SU}(2)\times {\rm SU}(2)$ or as the coset ${\rm SU}(2)^3/{\rm SU}(2)$, with the latter ${\rm SU}(2)$ diagonally embedded in ${\rm SU}(2)^3$. However, in this case, the approach of \cref{ssec:coset-reduction} leads to a solution with no gauge fields, and the final result is identical to the one in \cref{ssec:ads4X3Y3} with two internal spheres and equal fluxes, $n_X=n_Y$.

Note that with this example, each of the four homogeneous six-manifolds that admit a nearly-K\"{a}hler structure, $\mathbb{F}(1,2;3)$, $\C\mathbb{P}^3$, $S^6$, and $S^3\times S^3$, leads to a solution.

The remaining cosets, ${\rm SO}(7)/{\rm SO}(6)$, ${\rm SU}(4)/{\rm U}(3)$, and ${\rm SO}(5)/({\rm SO}(3)\times{\rm SO}(2))$, have no $H$ field and fall into the $k_M=0$ class of \cref{ssec:coset-reduction} (see \cref{eq:kM}).\footnote{${\rm SO}(7)/{\rm SO}(6)$ is the round sphere $S^6$, but the solution does not coincide with that of the previous section: the gauge field is in ${\rm SO}(6)$ rather than ${\rm SU}(3)$. ${\rm SU}(4)/{\rm U}(3)$ is topologically $\mathbb{CP}^3$, but with a Fubini--Study metric rather than the nearly-K\"ahler metric of \cref{ssub:cp3}, as well as a gauge group ${\rm U}(3)$ rather than ${\rm Sp}(1)\times {\rm U}(1)$.} These cases are vulnerable to higher-derivative corrections, as we explained after \cref{eq:kM=0-case}, and their ultimate fate is unclear. 

\subsection{Comments}
\label{ssec:comments}

The explicit examples that we have found with non-trivial gauge and three-form fluxes share two features: the internal homogeneous spaces admit nearly-K\"{a}hler structures, and the metrics are proportional to the identity.
In particular, while we explained the general Kaluza--Klein approach to the ${\rm SO}(16)\times\rm{SO}(16)$ string starting with the metric in \cref{eq:R2dij}, in \cref{ssec:ads-examples} we assumed the most general left-invariant metric on the coset; yet, for each example, we found the metric to take the form of \cref{eq:R2dij}.

It remains unclear whether either of the two features---the nearly-K\"{a}hler structure or the metric proportional to the identity---follows from the Einstein equations after replacing \cref{eq:R2dij} with the most general metric on $G/H$. 
Alternatively, this could be a consequence of taking $d=6$, which leaves only a limited number of choices for $G$ and $H$ that yield non-supersymmetric string vacua.

Another point remains unsettled. Our Kaluza--Klein approach differs from that of~\cite{Duff:1985cm}. They start from higher-dimensional equations (of the bosonic string) to derive lower-dimensional ones (of the supersymmetric heterotic string). In contrast, we have no higher-dimensional system to begin with. In fact, this explains the complexity of solving the Einstein equations separately in \cref{ssec:coset-reduction-einstein}. With a higher-dimensional system, possibly adding localized contributions, our approach would become completely systematic.
This would open up many more possibilities, fully exploiting the features of Kaluza--Klein reductions.

\section*{Acknowledgements}

We thank the organizers of the workshop \textit{Dark World to Swampland 2024}, where this work started.
This work was performed in part at the Aspen Center for Physics, which is supported by a grant from the Simons Foundation (1161654, Troyer). We thank M.~Garc\'{\i}a-Fern\'andez for an important discussion and C.~Angelantonj for useful comments.
SR is supported by the ERC Starting Grant QGuide101042568 - StG 2021.
AT is supported in part by the INFN, and by the MUR-PRIN contract 2022YZ5BA2. 

\appendix

\section{No scale separation}
\label{app:no_scale_separation}

The solutions of \cref{sec:3-form-vacua,sec:cpt-H-F} are not scale separated: $m_{\mathrm{KK}}\gg |\Lambda|^{1/2}$. We present here an argument against scale separation for ten-dimensional non-supersymmetric strings with a product metric and constant dilaton, assuming that $m_{\mathrm{KK}}$ scales with the internal curvature, $m_{\mathrm{KK}}^2\sim R_{(d)}$. Consider \cref{eq:constant-dilaton}, where $F_k$ can also denote gauge fields by adding traces. Taking partial traces of the Ricci tensor,
\begin{eq}\label[pluralequation]{eq:appendix_Ricci}
    R_{(10-d)}&=-(10-d)\sum_k \frac{k(\gamma+2)+\beta_k-\gamma}{8(2\gamma+5)}e^{(\beta_k+2)\phi_0}F_k^2\mcomma \\
    R_{(d)}&=\sum_k \frac{4k(2\gamma+5)-kd(\gamma+2)-d(\beta_k-\gamma)}{8(2\gamma+5)}e^{(\beta_k+2)\phi_0}F_k^2\mperiod
\end{eq}
For the types of internal spaces that we consider, the leading contribution to the internal trace $R_{(d)}$ comes from the space with the smallest radius: $R_{(d)}\sim R^{-2}$. The external trace scales with the AdS length: $R_{(10-d)}\sim L^{-2}$. Then, one can use the ratio of the two partial traces as a proxy for scale separation, as in~\cite{Gautason:2015tig}. From \cref{eq:appendix_Ricci}, we find
\begin{eq}
    \frac{R_{(d)}}{R_{(10-d)}}=\frac{\sum_k\left[d(\beta_k-\gamma)+kd(\gamma+2)-4k(2\gamma+5)\right]e^{2(\beta_k+2)\phi_0}F_k^2}{(10-d)\sum_k \left[\beta_k-\gamma+k(\gamma+2)\right]e^{2(\beta_k+2)\phi_0}F_k^2} \mperiod
\end{eq}
Since the combination $\beta_k-\gamma+k(\gamma+2)$ never vanishes for the ten-dimensional non-supersymmetric strings, the numerator and denominator are of the same order, and it is therefore impossible to achieve scale separation.

\bibliographystyle{utphys}
\bibliography{mybib}

\end{document}